\documentclass[aps,prd,amsmath,floats,twocolumn,superscriptaddress,nofootinbib,longbibliography]{revtex4-1}
\usepackage{bm}
\usepackage{mathrsfs}
\usepackage{fancyvrb}
\usepackage{amsmath}
\usepackage{amssymb}
\usepackage{graphicx}
\usepackage{accents}
\usepackage{algorithm}
\usepackage{algpseudocode}
\usepackage{setspace} 
\usepackage{color}
\usepackage{xcolor}
\usepackage{xspace}
\usepackage[normalem]{ulem}

\usepackage[colorlinks=true,linkcolor=blue,citecolor=blue]{hyperref}
\newcommand{\NRHybSur}{NRHybSur3dq8\xspace}

\newcommand{\UMassD}{\affiliation{Department of Mathematics and
    the Center for Scientific Computing \& Visualization Research,
    University of Massachusetts Dartmouth, Dartmouth, MA 02747, USA}}
\newcommand{\UMassDf}{\affiliation{Department of Physics, Department of Mathematics, and
    the Center for Scientific Computing \& Visualization Research,
    University of Massachusetts Dartmouth, Dartmouth, MA 02747, USA}}
\newcommand{\RIT}{\affiliation{Center for Computational Relativity and Gravitation, Rochester Institute of Technology, Rochester, New York 14623, USA}}
\newcommand{\Caltech}{\affiliation{Theoretical Astrophysics,
    California Institute of Technology, Pasadena, CA 91125, USA}}

\newcommand{\AEIp}{\affiliation{Max Planck Institute for Gravitational Physics
    (Albert Einstein Institute), Am M\"uhlenberg 1, 14476 Potsdam, Germany}} %
\newcommand{\Cornell}{\affiliation{Cornell Center for Astrophysics and Planetary Science, Cornell University, Ithaca, New York 14853, USA}}

\begin{document}

\title{Impact of subdominant modes on the interpretation of gravitational-wave signals from heavy binary black hole systems}

\author{Feroz H. Shaik}\UMassDf
\author{Jacob Lange}\RIT
\author{Scott E. Field} \UMassD
\author{Richard O'Shaughnessy}\RIT
\author{Vijay Varma} \Caltech
\author{Lawrence E. Kidder} \Cornell
\author{Harald P. Pfeiffer} \AEIp
\author{Daniel Wysocki}\RIT

\date{\today}

%%%%%%%%%%%%%%%%%%%%%%%%%%%%%%%%%%%%%%%%%%%%%%%%%%%%%%%%%%%%%%%
\begin{abstract}

Over the past year, a handful of new gravitational wave models have
been developed to include multiple harmonic modes thereby enabling for
the first time fully Bayesian inference studies including higher modes to be performed. 
Using one recently-developed numerical relativity surrogate model, NRHybSur3dq8, 
we investigate the importance of higher modes on parameter inference of
coalescing massive binary black holes. We focus on examples relevant to the current three-detector network 
of observatories, with a detector-frame mass set to $120 M_\odot$ and with signal
amplitude values that are consistent with plausible candidates for the next few observing runs.
We show that for such systems the higher mode content will be important for
interpreting coalescing binary black holes, reducing systematic bias, and computing properties of the remnant object.
Even for comparable-mass binaries and at low signal amplitude, the omission of higher modes
can influence posterior probability distributions. We discuss the impact of our results on 
source population inference and self-consistency tests of general relativity.
Our work can be used to better understand asymmetric binary black hole merger events, such as GW190412. 
Higher modes are critical for such systems, and their
omission usually produces substantial parameter biases.
\end{abstract}
\pacs{
Add some}
%  LIGO DCC:  https://dcc.ligo.org/P1900286
\maketitle
\section{Introduction} \label{sec:intro}

During their first and second observing runs, the Advanced LIGO \cite{2015CQGra..32g4001L} and Virgo \cite{gw-detectors-Virgo-original-preferred}  ground-based gravitational wave (GW) detectors have
identified several coalescing compact binaries
\cite{DiscoveryPaper,LIGO-O1-BBH,2017PhRvL.118v1101A,LIGO-GW170814,LIGO-GW170608,LIGO-GW170817-bns,LIGO-O2-Catalog}.  
GW detectors are exceptionally sensitive to very massive objects \cite{AstroPaper}, and the majority of
compact binaries observed to date are pairs of ${\cal O}(30M_\odot)$  binary black hole (BBH) systems \cite{LIGO-O2-Catalog}.   The early
analysis of these signals used semi-analytical approximations to 
general relativity \cite{gwastro-mergers-IMRPhenomP,2014PhRvD..89f1502T,2014PhRvD..89h4006P}. More recently, better approximations to general relativity have been developed
\cite{london2018first,Cotesta:2018fcv,varma2019surrogate,Blackman:2017dfb,PhysRevD.96.024058}, which include more of the available physics such as higher-harmonic modes.

Previous investigations have demonstrated that neglecting some of the physics present in real signals produces biased inferences for
compact binaries; conversely, including full physics enables sharper inferences.  
For instance, studies~\cite{Varma:2016dnf, bustillo2016impact, Capano:2013raa, Littenberg:2012uj,
Bustillo:2016gid, Brown:2012nn, Varma:2014, Graff:2015bba, Harry:2017weg} have
shown that the nonquadrupole modes, while being subdominant, can play a
nonnegligible role in detection and parameter estimation,
particularly for high signal-to-noise-ratio (SNR), large total mass, high mass
ratio, or systems favoring an edge-on orientation. In addition, nonquadrupole modes
can help break the degeneracy between the binary inclination and distance,
which is present for quadrupole-mode-only models (see
e.g.~\cite{london2018first, OShaughnessy:2014shr, Usman:2018imj, kumar2019constraining}).

The recent observation of GW190412 during the third observing run of LIGO and Virgo has highlighted the significance of higher-harmonic modes for the parameter estimation of unequal mass BBH mergers \cite{LIGOScientific:2020stg}. Using both precessing and aligned-spin models that included the effects of subdominant modes, it has been demonstrated that a measurable contribution of modes beyond the dominant quadrupolar mode was present in the data of GW190412. This underscores the need for such models for future observing runs.

For the first set of gravitational-wave observations, the massive binary black holes which dominate current observations produce short signals of modest SNRs. For the first event, GW150914~\cite{DiscoveryPaper}, where detailed followups were done, the systematic errors due to the
quadrupole-mode-only approximation is generally smaller than the statistical
errors~\cite{LIGO-O1-PENR-Systematics, Abbott:2016apu}, although higher modes may lead to
modest changes in some of the extrinsic parameter values~\cite{kumar2019constraining}.
A recent study~\cite{LIGO-O1-PENR-Systematics} considering GW150914-like events of near-equal mass and modest amplitude has concluded that neglecting sub-dominant waveform modes did not lead to bias and quadrupole-only models will suffice to characterize the observationally-accessible parameters of astrophysical binary black holes in the immediate future. However, as pointed out in Ref~\cite{LIGO-O1-PENR-Systematics}, at the time of that study there were no recovery models including higher modes and the systems considered were  $q \approx 1.2$ and a detector-frame total mass of 74. Recently, Chatzioannou 
et al.~\cite{chatziioannou2019properties} have reanalyzed GW170729, using IMRPhenomHM~\cite{london2018first},
SEOBNRv4HM~\cite{Cotesta:2018fcv}, and NRSur7dq2~\cite{PhysRevD.96.024058} and found that despite weak evidence for higher-order harmonic modes their inclusion in the analysis leads to increased support for unequal masses.

With newly developed multi-mode models it is now possible to revisit these questions.
We can now compute, for example, the true posteriors using recovery models with multiple harmonic modes that can then be compared to posteriors recovered with dominant modes only. Such comparisons will allow us to precisely quantify the information gained by using subdominant modes. 
For example, even for an equal-mass system, we observe that the posterior produced without subdominant modes will experience a noticeable shift towards (incorrectly) favoring lighter binary systems with more negative
$\chi_{\rm eff}$ values (cf.~Figures~\ref{fig:q1} and~\ref{fig:SNR1}). In fact this preferential bias appears to be a common feature across many of the cases we have considered.

In this paper, we use concrete examples of end-to-end parameter inference to quantify
how much approximations that neglect subdominant modes can 
impact the interpretation of gravitational-wave events.
Unlike previous studies, which typically used either a single detector, low signal-to-noise ratios (SNRs), or a Fisher matrix analysis, our fully Bayesian study uses a three-detector network with SNRs typical of detections expected in the near future. We demonstrate these inference biases occur even at moderate signal amplitude for some configurations, growing
extreme at amplitudes expected for some sources when LIGO reaches design sensitivity~\cite{Abbott:2016wya}. 

We also explore additional physics that can be extracted with non-quadropoles modes using a spin-aligned model, such as improved measurability of individual spin components, final mass and spin properties of the remnant, black hole kicks~\cite{bustillo2018tracking}, source population inference, and self-consistency tests of general relativity. For example, in the context of non-spinning BBH systems, Ref.~\cite{2018PhRvD..98b4019P} has demonstrated that when higher-modes are omitted from the recovery model, its effect can mimic deviations from General Relativity.

Our examples target sources with detector-frame masses $M_z\simeq 120 M_\odot$, comparable to the detector-frame masses
expected for typical near-future binary black hole observations (e.g., pairs of $35 M_\odot$ BHs at moderate redshift).  For comparison,
as ground-based detector networks approach design sensitivity and regularly detect sources near $z\simeq 1$, a merging
pair of BHs near the pair-instability mass-gap  ($50 M_\odot$) observed at $z\simeq 1$ would have a detector-frame mass of $M_z\simeq 200
M_\odot$~\cite{belczynski2016effect}.
We also consider target sources with mass ratios in the range $1 \le q \le 7$.
To date most LIGO/Virgo events show support only for systems with mass ratios less than 2~\cite{LIGO-O2-Catalog}. The recent observation of GW190412 \cite{LIGOScientific:2020stg} has now shown that we should expect to observe larger mass ratio systems in the future. For example, 
unequal mass systems are generically expected for BBH mergers within the accretion disks of active galactic nuclei~\cite{mckernan2019montecarlo}. Furthermore, the first and second observing runs~\cite{LIGO-O2-Catalog} have already observed compact objects over a mass
range of $1.3 M_{\odot}$ to $85 M_{\odot}$ suggesting combinations involving
mass-ratios as large as $7$ are not unreasonable for LIGO/Virgo to observe. 

This paper is organized as follows.  
In Section  \ref{sec:preliminaries} we introduce the GW signal model and parameter inference techniques
used in this work. 
In Section \ref{sec:bias} we survey the results of parameter inference on a sequence of synthetic high-mass binary black
holes with systematically-varied mass ratio, spin, and signal amplitude.   We specifically address how higher modes
impact inference, comparing parameter inferences  performed with the
full NRHybSur3dq8 model and with a model truncated to include only $\ell =2$ modes.
In Section \ref{sec:discussion} we discuss some consequences of our analysis. 
We conclude in Section \ref{sec:conclude} with some brief remarks and future work.

\section{Preliminaries} \label{sec:preliminaries}

\subsection{Gravitational Wave Model} 
\label{sec:GW_model}

A coalescing compact binary in a quasicircular orbit can be completely characterized by eight intrinsic
parameters, namely the individual masses, $m_i$, and spin vectors, $\mathbf{S}_i$, of each compact object. 
Gravitational waveform models and 
inference codes often employ parameterizations involving the system's total mass, $M=m_1+m_2$, the mass ratio,
\begin{equation}
\label{eq:q}
q=m_{1}/m_{2} \,,
\end{equation}
where $m_1 \geq m_2$, the dimensionless spins, 
\begin{equation}
\label{eq:chii}
\bm{\chi_{i}}=\bm{S_{i}}/m_{i}^{2} \,,
\end{equation}
on the individual black holes (BHs), and an 
effective spin parameter~\cite{damour2001coalescence,racine2008analysis,ajith2011inspiral},
\begin{equation}
\label{eq:chieff}
\chi_{\rm eff}=(\bm{S_{1}}/m_{1}+\bm{S_{2}}/m_{2})\cdot\hat{L}/M \,,
\end{equation}
which is a weighted combination of the spins projected along the normalized orbital angular momentum vector $\hat{L}$. 
We will express the dimensionless spins in terms of Cartesian components $\chi_{i,x},\chi_{i,y}, \chi_{i,z}$, expressed
relative to the source frame. We define this frame such that the 
$z-$axis is along the orbital angular momentum direction, which is constant for nonprecessing BBH systems. Since our focus is on the impact of higher-harmonic 
modes, we restrict ourselves to the 4-dimensional space of nonprecessing BBHs where non-quadropole, inspiral-merger-ringdown (IMR) models are more mature. Such systems are characterized by $\chi_{i,x} = \chi_{i,y} = 0$ and $|\chi_{1z}|,|\chi_{2z}|\leq 1$.

When discussing waveform models, it is common practice to introduce a complex gravitational-wave strain
\begin{align} \label{eq:waveform_modes}
h_{+}(t;t_c,\iota,\phi_c,\vec{\lambda}) &- {\mathrm i} h_{\times}(t;t_c,\iota,\phi_c,\vec{\lambda}) \nonumber \\
& =  \sum_{\ell=2}^{\infty} \sum_{m=-\ell}^{\ell} h^{\ell m}(t-t_c;\vec{\lambda}) {}_{-2}Y_{\ell m} (\iota,\phi_c,) \, ,
\end{align}
which is subsequently decomposed into a basis of spin-weighted spherical harmonics ${}_{-2}Y_{\ell m}$.
Here $\vec{\lambda} \equiv (q,M,\chi_{1z},\chi_{2z})$ is used to
denote the signal's dependence on the intrinsic parameters, 
$\iota$ is the inclination angle between the orbital angular momentum of the binary and line-of-sight to the detector,
$t_c$ is the coalescence time, and 
$\phi_c$ is the orbital phase at coalescence. Most gravitational waveform 
models make predictions for the modes $h^{\ell m}(t)$, from which the gravitational-wave strain 
detected by a ground-based interferometer,
\begin{align} \label{eq:strain}
h(t;\vec{\Lambda}) = 
& \frac{1}{r} F_{+}     \left(\text{ra},\text{dec},\psi\right) h_{+}     (t;t_c, \iota,\phi_c,\vec{\lambda}) + \nonumber \\
& \frac{1}{r} F_{\times}\left(\text{ra},\text{dec},\psi\right) h_{\times}(t;t_c, \iota,\phi_c,\vec{\lambda}) \, ,
\end{align}
is readily assembled.
The signal's dependence on four
additional extrinsic parameters are the polarization angle ($\psi$),
the luminosity distance to the source's center-of-mass ($r$),
and sky location determined by the right ascension ($\text{ra}$) and declination ($\text{dec}$). 
The antenna patterns $F_{(+,\times)}$ project the GW's 
$+$- and $\times$-polarization states, $h_{(+,\times)}$, into the detector's frame. 
We shall use $\vec{\Lambda} \equiv (\text{ra},\text{dec},\psi,r,t_c,\iota,\phi_c, \vec{\lambda} )$
to denote the signal's dependence on all 11 parameters defining the problem. 

Until recently, all spinning
IMR models had set $h^{\ell m} = 0$ except for the dominant $h^{2, \pm2}$ quadrupole modes.
The expectation had been that higher modes won't substantially affect parameter inference
for the O2 gravitational-wave observations, which are characterized by low SNRs and mostly face-on events of near-equal mass~\cite{Littenberg:2012uj,Varma:2014,LIGO-O1-PENR-Systematics}.

Over the past year or so, three new aligned-spin IMR models have been built to include non-quadropole modes: (i) a phenomenological frequency-domain model, IMRPhenomHM~\cite{london2018first}, includes the $(\ell, | m | ) = (2,2), (3,3), (4,4), (2,1), (3,2), (4,3)$ modes; (ii) an effective-one-body time-domain model, SEOBNRv4HM~\cite{Cotesta:2018fcv}, includes a similar set of $(\ell, | m | ) = (2,2), (3,3), (4,4), (5,5), (2,1)$ modes; (iii) a time-domain surrogate model for hybridized nonprecessing numerical relativity waveforms, NRHybSur3dq8~\cite{varma2019surrogate}, includes all of the $\ell \leq 4$ and $(5,5)$ spin-weighted spherical harmonic modes but not the $(4,1)$ or $(4,0)$ modes. 

Our study will use NRHybSur3dq8 as it both includes the most modes and is expected to be more accurate when evaluated within its training region (cf. Fig 6 from Ref.~\cite{varma2019surrogate}) of mass ratio $q\leq8$, and
$|\chi_{1z}|,|\chi_{2z}|\leq0.8$. For the $20~\text{Hz}$ starting frequency considered here, this model is valid for the entire LIGO band for stellar mass binaries with total masses as low as $2.25\,M_{\odot}$. 
We evaluate the model through the Python package GWSurrogate~\footnote{We use GWsurrogate version 0.9.\{4,5\}, which exactly agrees with the lalsimulation~\cite{LAL} implementation of the NRHybSur3dq8 model.}~\cite{gwsurrogate,Field:2013cfa}. The GWSurrogate package 
provides direct access to the GW's harmonic modes $h^{\ell m}(t)$ appearing in the sum~\eqref{eq:waveform_modes}.

By comparing to NR, Ref.~\cite{varma2019surrogate} has computed the NRHybSur3dq8 model's mismatches (averaged over many points on the sky) as a function of total mass using the Advanced LIGO design sensitivity noise curve. For the $120 M_{\odot}$ total mass systems predominantly used in our studies, the single-detector mismatches have a median value of $1\times 10^{-5}$.  A sufficient condition for two waveform models (in this case NR and NRHybSur3dq8) to be considered 
indistinguishable is~\cite{Flanagan1998a, Lindblom2008, McWilliams2010b,LIGO-O1-PENR-Systematics}
\begin{align}
\label{eq:indistinguishable}
\mathcal{M} < \frac{D}{2\rho^2} \,,
\end{align}
where $\mathcal{M}$ is the mismatch and $\rho$ is the signal-to-noise ratio (SNR). Here $D$ is an unknown constant that is sometimes associated with the number of model parameters~\cite{PhysRevLett.118.051101}, with $D = 4$ for our spinning BBH model. 
Furthermore, if the likelihood can be approximated by a Gaussian then an expression for $D$ can be obtained in terms of 
a chi-squared distribution with $4$ degrees of freedom~\cite{baird2013degeneracy}. Using this value for $D$ and a typical mismatch value quoted above, we find that the NRHybSur3dq8 model will give robust parameter estimates so long as $\rho \lesssim 450$. Even using pessimistic values ($D=1$ and the 95th percentile of mismatch errors $7\times10^{-5}$) we find that NR and our model will be indistinguishable according to Eq.~\eqref{eq:indistinguishable} so long as $\rho \lesssim 85$. 

For context, we note that in the first and second observing runs most BBH signals had a network SNR of about $15$ and
spanning a range of $10$ to $30$. In the upcoming observing run we would expect typical BBH SNRs to be between $10$ and
roughly 40, based on the cumulative distribution of the loudest SNR $\rho$  among $n$ identified events
($[1-(\rho/10)^3]^n$ using a fiducial value  $n=30$).
We caution the reader that in practice the
condition in Eq.~\eqref{eq:indistinguishable} should only be taken as a rough estimate. For instance, it features an unknown
constant $D$ while the NR waveforms themselves have small, systematic sources of error that would prevent any model to claim indistinguishability from general relativity beyond estimates of this systematic error~\cite{boyle2019sxs}. Finally, the definition of ``indistinguishable" is not synonymous with
``identical posterior distributions". Indeed, Fig.~\ref{fig:SNR1} shows that even for simple systems at low SNR, which 
easily satisfy Eq.~\eqref{eq:indistinguishable}, there can be noticeable discrepancies between the recovered posteriors. 
For example, using a single interferometer the mismatch between $\ell_{\rm max}=5$ and $\ell_{\rm max}=2$ models for a non-spinning, equal-mass
system is 0.0021, and so Eq.~\eqref{eq:indistinguishable} is satisfied at SNRs less than 30.

Due to the absence of higher-mode models for spinning BBH systems until recently, previous parameter-inference studies
that have focused on the information content available higher modes have 
either used quadrupole-only (recovery) models or leveraged the Fisher matrix framework. For high-accuracy,
high-SNR scenarios involving the 3-detector network neither of these are fully sufficient. For example, with the quadrupole-only
model the reference (``true") posterior will not be possible to compute in principle. Additionally, some of these models may
have modeling errors in the dominant mode that could become noticeable at high SNR
\cite{2013PhRvD..87j2002V,2017PhRvD..96l4041W,2014PhRvL.112j1101F}. 

\subsection{Bayesian Inference}

The likelihood of GW data in Gaussian noise has the form  (up to normalization),
\begin{equation}
\label{eq:lnL}
\ln {\cal L}(\bm{\lambda} ,\theta )=-\frac{1}{2}\sum\limits_{k}\langle h_{k}(\bm{\lambda} ,\theta )-d_{k} |h_{k}(\bm{\lambda} ,\theta )-d_{k}\rangle _{k}-\langle d_{k}|d_{k}\rangle _{k},
\end{equation}
where $h_{k}$ are the predicted response of the k$^{th}$ detector due to a source with parameters ($\bm{\lambda}$, $\theta$) and
$d_{k}$ are the detector data in the k$^{th}$ instrument; $\bm{\lambda}$ denotes the combination of redshifted total mass $M_{z}$ and the
remaining intrinsic parameters needed to uniquely specify the binary's dynamics; $\theta$ represents the
seven extrinsic parameters (4 spacetime coordinates for the coalescence event and 3 Euler angles for the binary's
orientation relative to the Earth); and $\langle a|b\rangle_{k}\equiv
\int_{-\infty}^{\infty}2df\tilde{a}(f)^{*}\tilde{b}(f)/S_{h,k}(|f|)$ is an inner product implied by the k$^{th}$ detector's noise power 
spectral density (PSD) $S_{h,k}(f)$. 
%In all calculations, we adopt the fiducial O1 noise power spectra associated with data near GW150914 \cite{DiscoveryPaper}.
In practice we adopt both  low- and high- frequency cutoffs $f_{\rm max},f_{\rm min}$ so all inner products are modified to
\begin{equation}
\label{eq:overlap}
\langle a|b\rangle_{k}\equiv 2 \int_{|f|>f_{\rm min},|f|<f_{\rm max}}df\frac{[\tilde{a}(f)]^{*}\tilde{b}(f)}{S_{h,k}(|f|)}.
\end{equation}
The joint posterior probability of $\bm{\lambda} ,\theta$ follows from Bayes' theorem:
\begin{equation}
p_{\rm post}(\bm{\lambda} ,\theta)=\frac{ {\cal L}(\bm{\lambda} ,\theta)p(\theta)p(\bm{\lambda})}{\int d\bm{\lambda} d\theta {\cal L}(\bm{\lambda} ,\theta)p(\bm{\lambda})p(\theta)},
\end{equation}
where $p(\theta)$ and $p(\bm{\lambda})$ are priors on the (independent) variables $\theta ,\bm{\lambda}$.  
Following most previous work \cite{Veitch:2015,gwastro-PENR-RIFT,LIGO-O2-Catalog}, we adopt uninformed separable priors for parameter inference.
\subsection{RIFT}

To construct the posterior distribution, we use the RIFT algorithm \cite{gwastro-PENR-RIFT}, which iteratively constructs and refines an
approximation to the marginal likelihood
\begin{equation} \label{eq:lnLmarg}
 {\cal L}_{\rm marg}\equiv\int  {\cal L}(\bm{\lambda} ,\theta )p(\theta )d\theta \,,
\end{equation}
which appears in Bayes' theorem for the marginal posterior distribution for $\bm{\lambda}$.  We use an existing program (ILE,
which Integrates the Likelihood over Extrinsic parameters) to perform the necessary marginalization, for each fixed
source \cite{gwastro-PE-AlternativeArchitectures,Abbott:2016apu,2017PhRvD..96j4041L,2017CQGra..34n4002O},
by marginalizing the likelihood of the data over the seven parameters characterizing the spacetime coordinates and
orientation of the binary relative to the earth; see \cite{gwastro-PE-AlternativeArchitectures,gwastro-PENR-RIFT-GPU} paper for technical details.

To achieve rapid turnaround times, we use the new
GPU-accelerated implementation of ILE \cite{gwastro-PENR-RIFT-GPU}. Working on the CARNiE cluster, which includes 15 NVIDIA Tesla V100 GPU-enabled nodes, our current configuration completes each of the binary black hole analyses
presented in this work in about 15 to 20 hours. When using all 15 GPUs, a single ILE step for an SNR=30 case takes about 1 hour to finish.

% Interpolation
Following the RIFT algorithm \cite{gwastro-PENR-RIFT}, we iteratively construct an approximation to the likelihood by generating and drawing from
approximate posterior distributions, until our posterior distribution converges.   At each iteration, the  likelihood is
approximated using Gaussian process regression with a  squared-exponential kernel, with hyperparameters tuned to the
likelihood evaluations available at that iteration.

\section{Intrinsic-parameter biases}
\label{sec:bias}
In this section, we present parameter estimation (PE) results from sources listed in Table \ref{tab:SourceParameters}. 
All synthetic datasets use PSDs generated from data near GW170814~\cite{LIGO-GW170814}, when all three detectors were operational, and are created with zero noise realizations. Specifically the synthetic detector data is exactly equal to the expected response due to our GW source.
Since detector noise is assumed to be colored Gaussian noise with zero mean, using zero noise with the likelihood defined in Eq.~\eqref{eq:lnL}
makes our analysis equivalent to an average over an ensemble of analyses which use infinitely many noise realizations~\cite{LIGO-O1-PENR-Systematics}. For all runs, $f_{\rm min}$ and $f_{\rm max}$ from Eq. \ref{eq:overlap} are $20$ Hz and $2000$ Hz, respectively.

Each synthetic dataset includes an injected signal from the expected response at each detector due to our GW source
using the \NRHybSur{} model and including all of the surrogate's available $\ell_{\rm max}=5$ modes (see Sec.~\ref{sec:GW_model} for the exact modes, which, for example, only includes $(5,5)$ among the $\ell=5$ modes).
The model generates a waveform such that the 
instantaneous initial frequency of the $(2, 2)$ mode has a frequency of $8~\text{Hz}$,
which ensures the $(5,5)$ mode's instantaneous initial frequency is out-of-band. We taper 
the beginning and end portions of the waveform to avoid artificial oscillations in the Fourier
domain. In particular, since NR waveforms (and therefore the  \NRHybSur{} model)
do not go to zero by the end of the simulation, 
we have found it necessary to taper the last 
%$40$M (in dimensionless units) 
portion of the ringdown signal.

We adopt conventional mass and distance priors, uniform in
detector-frame mass and in the cube of the luminosity distance. For our nonprecessing spins, we adopt a uniform prior for $\chi_{i,z} \in[-0.9,0.9]$. Sec.~\ref{sec:spin_priors} considers the effect of using an alternative spin prior in the context of high SNR events.

Each of the following subsections describe a set of related runs, varying one of the problems' parameters at a time. For each source configuration, we present parameter estimates recovered using all of
the available higher modes $\ell_{\rm max}=5$ (we may sometimes refer to this as the ``true" or reference posterior) and compare with posteriors recovered using the same model restricted to only the $\ell_{\rm max}=2$ modes (using $|m|=\{2,1\}$). In subsections \ref{subsec:q1} ($q=1$), \ref{subsec:q4} ($q=4$), 
and \ref{subsec:q7} ($q=7$) we vary the spin configurations 
of $\chi_{\rm 1z} = \chi_{\rm 2z}=\{-0.8,-0.5,0.0,0.5,0.8\}$ while keeping the network SNR fixed at $30$~\footnote{Given a fixed starting frequency, systems with their BH component spins (anti-)aligned with the orbital angular momentum will be (shorter) longer. As a result, to achieve a fixed SNR the spin (anti-)aligned systems must be place located (closer) farther as compared to a reference non-spinning system.}. For this sequence of runs, our choice of inclination angle, $\iota = 3\pi/4$, is neither face-on nor edge-on, but rather constitutes a ``general" configuration. In subsection~\ref{subsec:SNRs} we consider varying the SNR to explore its effect on marginalized posterior distributions.

It is known that the contribution of subdominant modes 
towards the signal's power increases as the inclination angle is increased from a face-on ($\iota = 0$) to an edge-on ($\iota = \pi/2$) configuration. As such, we expect our observed biases to be larger (smaller) when compared to a face-on (edge-on) system at the same network SNR value. This general expectation was recently confirmed by Kalaghatgi et al.~\cite{Kalaghatgi:2019log}, where the importance of subdominant modes for non-spinning systems was quantified by systematically varying the inclination angle across a range of values. In our study we have instead fixed the inclination angle to a value typical of an O2 event~\cite{LIGO-O2-Catalog} while systematically exploring the impact due to spin. As such our results are complementary to those of Ref.~\cite{Kalaghatgi:2019log}.

\begin{table}
\begin{ruledtabular}
\begin{tabular}{lcccccccccc}
    ID\# & $\iota$ & $q$ & $M$ ($M_{\odot}$) & $\chi_{\rm 1z}$ & $\chi_{\rm 2z}$ & SNR\\ \hline
    1 & $\pi/4$ & 2.267 & 127.1 & 0.72 & 0.0 & 30\\
    2 & $3\pi/4$ & 1.00 & 120.0 & -0.80 & -0.80 & 30\\
    3 & $3\pi/4$ & 1.00 & 120.0 & -0.50 & -0.50 & 30\\
    4 & $3\pi/4$ & 1.00 & 120.0 & 0.0 & 0.0 & 10,30,70\\
    5 & $3\pi/4$ & 1.00 & 120.0 & 0.50 & 0.50 & 30\\
    6 & $3\pi/4$ & 1.00 & 120.0 & 0.80 & 0.80 & 30\\
    7 & $3\pi/4$ & 4.00 & 120.0 & -0.8 & -0.8 & 30\\
    8 & $3\pi/4$ & 4.00 & 120.0 & -0.5 & -0.5 & 10,30,70\\
    9 & $3\pi/4$ & 4.00 & 120.0 & 0.0 & 0.0 & 30\\
    10 & $3\pi/4$ & 4.00 & 120.0 & 0.5 & 0.5 & 30\\
    11 & $3\pi/4$ & 4.00 & 120.0 & 0.8 & 0.8 & 30\\
    12 & $3\pi/4$ & 7.00 & 120.0 & -0.8 & -0.8 & 30\\
    13 & $3\pi/4$ & 7.00 & 120.0 & -0.5 & -0.5 & 30\\
    14 & $3\pi/4$ & 7.00 & 120.0 & 0.0 & 0.0 & 30\\
    15 & $3\pi/4$ & 7.00 & 120.0 & 0.5 & 0.5 & 30\\
    16 & $3\pi/4$ & 7.00 & 120.0 & 0.8 & 0.8 & 30\\
    %17 & $3\pi/4$ & 4.00 & 120.0 & 0.0 & 0.0 & 10\\
    %18 & $3\pi/4$ & 4.00 & 120.0 & 0.0 & 0.0 & 20\\
    %19 & $3\pi/4$ & 4.00 & 120.0 & 0.0 & 0.0 & 50\\
    %20 & $3\pi/4$ & 4.00 & 120.0 & 0.0 & 0.0 & 70\\
%    21 & $3\pi/4$ & 2.00 & 50.0 & 0.0 & 0.0 & 0.0 & 0.0 & 0.0 & -0.8 & 30\\
\end{tabular}
\caption{\label{tab:SourceParameters}\textbf{Parameters of synthetic sources}:  This table shows the parameters of all
  the synthetic sources used in this paper. $\iota$ is the inclination angle between the line of sight of the observer and the total angular momentum vector, $q$ is the mass
  ratio defined with $q>1$ (see Eq. \ref{eq:q}), $M$ is the detector-frame total mass, and $\chi_{*}$ are the components of the
  normalized spins (see Eq. \ref{eq:chii}). As we use a non-precessing model, we set all of the in-plane spin components to $0$.
  All luminosity distances are set such that the network signal-to-noise ratio achieves the value specified under the SNR column. For example, in our $q=7$ sequence the most extreme values of spin, $\chi_{\rm eff} = -0.8$ and $\chi_{\rm eff} = 0.8$, are located at $181.4720$ Mpc and $452.5185$ Mpc, respectively. This large discrepancy in distance is due to the orbital hangup effect and is explained in greater detail in Fig.~\ref{fig:mode_power}. Other extrinsic parameters are fixed to the following values: right ascension is RA=0.0, declination is DEC=1.5707963, and the polarization angle is $\psi=\pi/4$.
}
\end{ruledtabular}
\end{table}

\subsection{q=1}
\label{subsec:q1}
\begin{figure*}
\includegraphics[width=\textwidth,height=\textwidth]{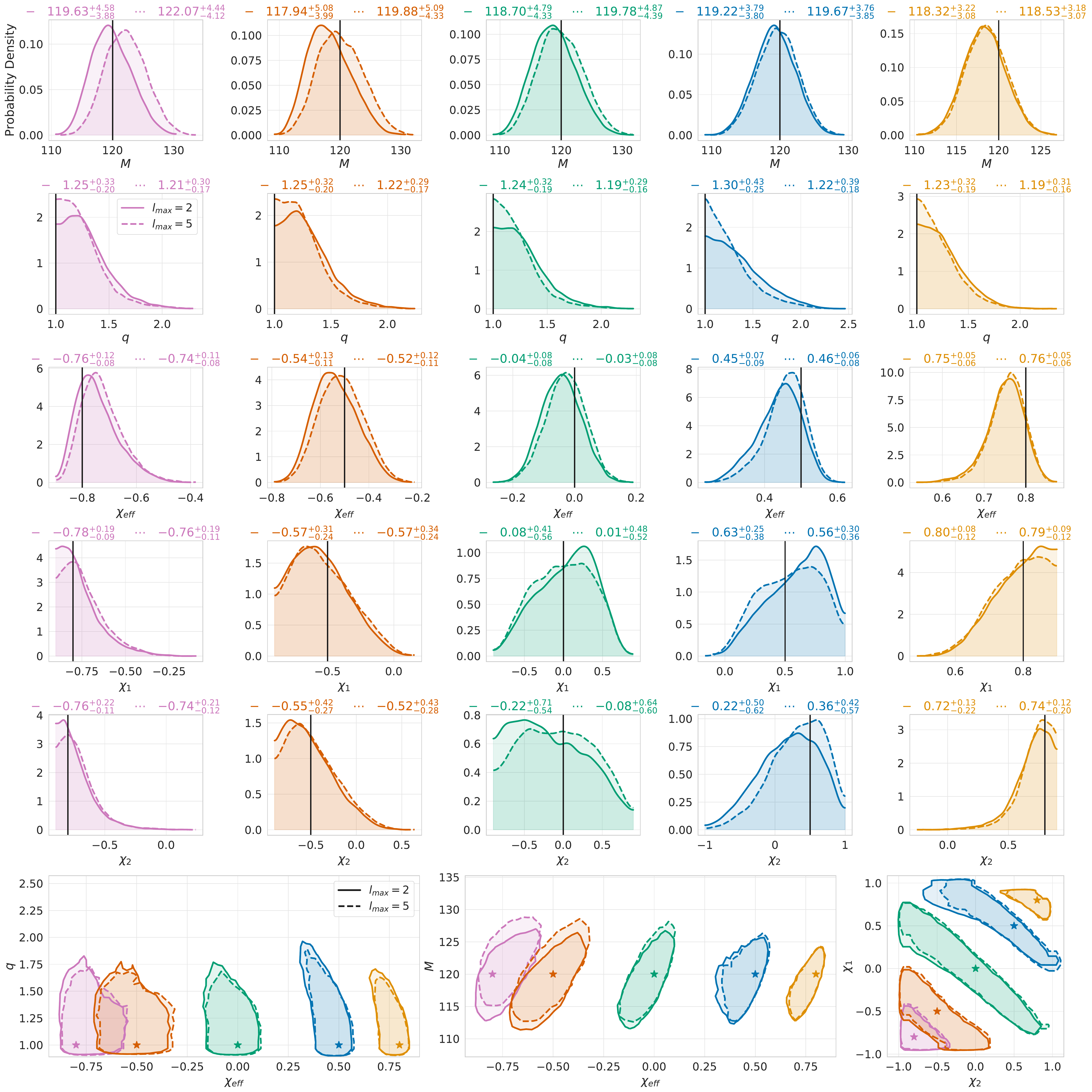}
\caption{\label{fig:q1} \textbf{Non-HM and HM runs for q=1 spin set, with SNR=30 and M=120}: The first five rows 
show the $M, q, \chi_{\rm eff}, \chi_{1z}, \chi_{2z}$ one-dimensional marginal distributions, where among this 
set of figures each column corresponds to a different synthetic source 
recovered with either all $\ell_{\rm max}=5$ modes (dashed line) or $\ell_{\rm max}=2$ modes (solid line). Our figures are organized such that the injected spin is systematically increased from left to right, where
the synthetic source runs 
are ID2 ($\chi_{\rm eff} = -.8$), ID3 ($\chi_{\rm eff} = -.5$), ID4 ($\chi_{\rm eff} = 0$), ID5 ($\chi_{\rm eff} = .5$), and ID6 ($\chi_{\rm eff} = .8$). In each figure's title, we report the median value and the 90\% confidence intervals of the marginalized 1D distribution for the $\ell_{\rm max}=2$ (left) and $\ell_{\rm max}=5$ (right) cases. A solid black vertical line denotes the true parameter value. The final bottom row corresponds to the joint distributions for $q$ vs $\chi_{\rm eff}$, M vs $\chi_{\rm eff}$, and $\chi_{1,z}$ vs $\chi_{2,z}$ for all five injections.
}
\end{figure*}

We first look at a set of equal mass runs with the different spin configurations mentioned above. It is well known that the relative power of subdominant harmonic modes are minimized for equal mass BBH systems, so these cases are expected to minimize bias. Previous studies~\cite{Littenberg:2012uj,Varma:2014,Varma:2016dnf,LIGO-O1-PENR-Systematics} have either found negligible bias (for face-on systems), small bias (for edge-on systems), or quoted results averaged over the source orientation where again only very small biases were found. At the time of these studies~\cite{Littenberg:2012uj,Varma:2014,Varma:2016dnf,LIGO-O1-PENR-Systematics}, however, there were no recovery models for near-equal mass spinning BBH systems including subdominant modes so these results were only suggestive. Here we confirm the general expectation of smaller bias at $q=1$, while also making more precise the nature of the bias by comparing the true posterior to the approximate one found with $\ell_{\rm max} = 2$ modes only. For example, in all cases the
true posterior's peak is located at $q=1$, while some of the biased posteriors have a non-negligible offset often peaking closer to $q \sim 1.25$. From Fig.~\ref{fig:q1} we also observe noticeable shifts in the posteriors 90\% confidence region for anti-aligned configurations.

Figure \ref{fig:q1} shows
the posterior distributions of the intrinsic parameters
for all the different spin
configurations. The solid lines represent runs that were done with $\ell_{\rm max}=2$ modes, and the dashed lines represent runs
that include all available $\ell_{\rm max}=5$. For each run, there is some degree of difference between the $\ell_{\rm max}=2$
and $\ell_{\rm max}=5$ runs. As anticipated by Ref.~\cite{Varma:2016dnf}, which used a non-Bayesian approach and a single detector, this discrepancy between the two distributions become more
extreme as the spins increase toward negative spin. For example, for negative spins there are noticeable shifts in the $M$ vs $\chi_{\rm eff}$ posteriors. We emphasize that even for the simplest case (equal mass and zero
spin), differences between the two results are visible. Although parameter recovery is not biased in the sense that all of the injection values lie within their 90\% confidence regions, it is also clear from the figure that the median recovered using all subdominant modes is almost always closer to the injection value.
This is contrary to
the general expectation that subdominant modes are largely irrelevant for equal-mass systems
\cite{Littenberg:2012uj,Varma:2014,LIGO-O1-PENR-Systematics}. Section \ref{subsec:SNRs} explores how different network SNRs affect the bias for these systems; Appendix \ref{app:zerospin} follows up on the curious differences seen in the simplest case of zero spin, equal mass.

\subsection{q=4}
\label{subsec:q4}

We next increase our set of sources to $q=4$, a configuration that is most relevant to GW190412-like events. Similar to the $q=1$ case, as far as we are aware, the existing literature for parameter estimation is comprised of results for non-spinning recovery models~\cite{Littenberg:2012uj}, results for near-equal mass without multi-mode recovery
models~\cite{LIGO-O1-PENR-Systematics}, or Fisher matrix-based studies~\cite{Varma:2014,Varma:2016dnf}. None of those studies
consider the 3-detector network configuration and a multi-modal recovery model with fully Bayesian inference. At larger
mass ratios, our study confirms the general expectations described in Ref.~\cite{Varma:2016dnf}, although the observed
bias is often even larger than expected; compare to the typical errors indicated by
corresponding green, red, and blue curves in Figure 6
of  Ref.~\cite{Varma:2016dnf} for our fiducial mass. 
We also are able to more carefully quantify the nature of the bias by comparing to the true posteriors.
In particular, similar to the $q=1$ systems just considered, neglecting subdominant modes consistently shifts the posterior towards more extreme anti-aligned spin configurations with lighter total mass.

\begin{figure*}
\includegraphics[width=1.0\textwidth]{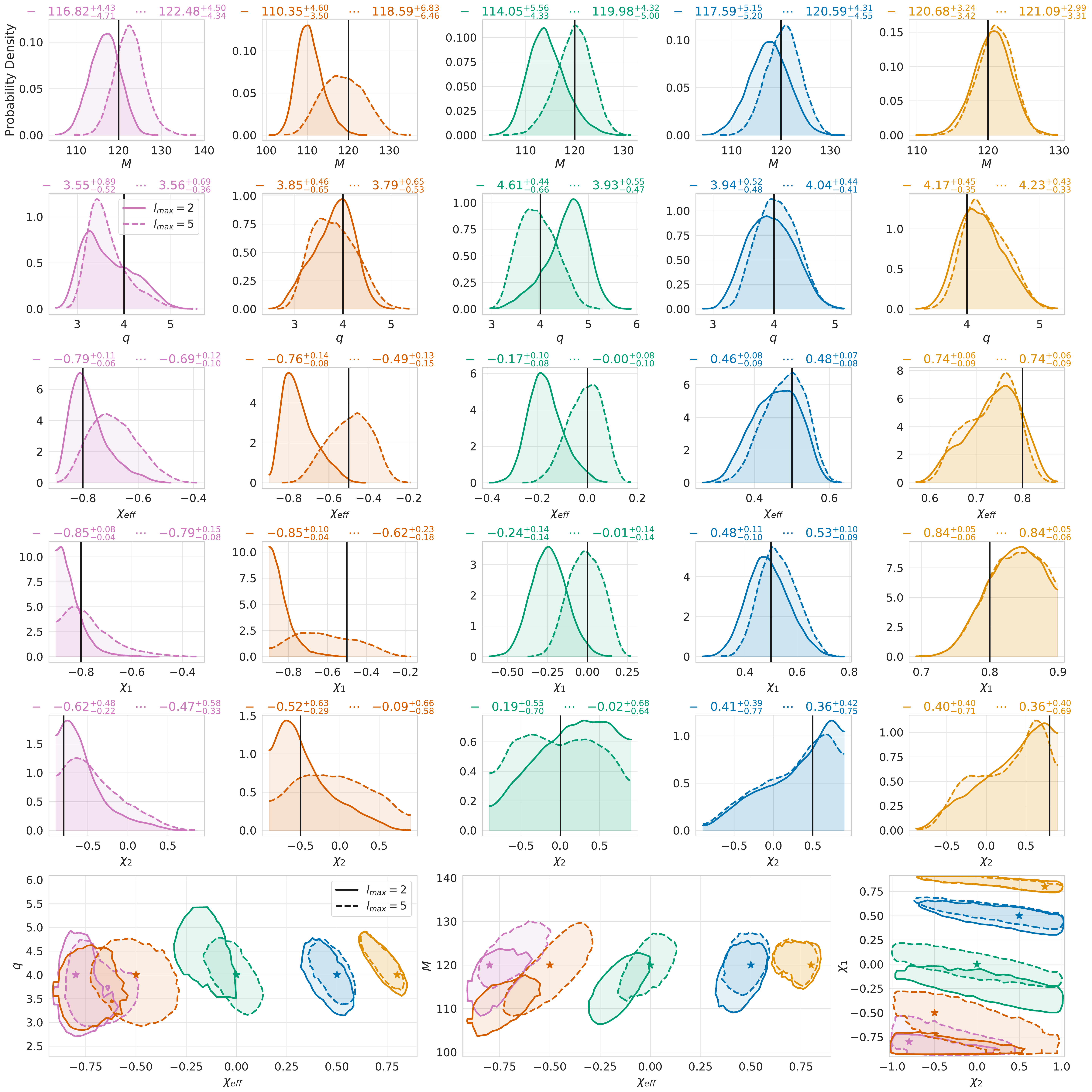}
\caption{\label{fig:q4} \textbf{Non-HM and HM runs for q=4 spin set, with SNR=30 and M=120}: The first five rows 
show the $M, q, \chi_{\rm eff}, \chi_{1z}, \chi_{2z}$ one-dimensional marginal distributions, where among this 
set of figures each column corresponds to a different synthetic source 
recovered with either all $\ell_{\rm max}=5$ modes (dashed line) or $\ell_{\rm max}=2$ modes (solid line). Our figures are organized such that the injected spin is systematically increased from left to right, where
the synthetic source runs 
are ID7 ($\chi_{\rm eff} = -.8$), ID8 ($\chi_{\rm eff} = -.5$), ID9 ($\chi_{\rm eff} = 0$), ID10 ($\chi_{\rm eff} = .5$), and ID11 ($\chi_{\rm eff} = .8$). In each figure's title, we report the median value and the 90\% confidence intervals of the marginalized 1D distribution for the $\ell_{\rm max}=2$ (left) and $\ell_{\rm max}=5$ (right) cases. A solid black vertical line denotes the true parameter value. The final bottom row corresponds to the joint distributions for $q$ vs $\chi_{\rm eff}$, M vs $\chi_{\rm eff}$, and $\chi_{1,z}$ vs $\chi_{2,z}$ for all five injections.
}
\end{figure*}

Figure \ref{fig:q4} shows the posterior distributions for $\chi_{\rm eff}$ vs $q$ and $\chi_{\rm eff}$ vs $M$ for all the different spin configurations. The solid lines again represent runs that were done with $\ell_{\rm max}=2$ modes, and the dashed lines represent runs that include all available $\ell_{\rm max}=5$. Similar to Section \ref{subsec:q1}, we again see that the differences become more extreme as the spins increase toward negative spin. Comparing the same spin configures between $q=1$ and $q=4$ runs, it is clear that increasing the mass ratio dramatically increases the bias between the non-HM and HM runs. In particular, there are now many cases where parameter estimates recovered with $\ell_{\rm max}=2$ modes do not lie within their 90\% confidence regions. Looking at the two-dimensional posteriors, for example, shows many cases where either the $\ell_{\rm max}=2$ posterior either does not contain the injection value or it is noticeably shifted from the true posterior. 
By comparison, in almost all of the $\ell_{\rm max}=5$ cases, the marginal posteriors almost perfectly peak at the true parameters. One notable exception is the $\chi_{\rm eff}=-0.8$ case (the purple distributions in Figure \ref{fig:q4}) where the true parameters seem to lie just inside the 90\% confidence region. We suspect this is due to a combination of (i) the injection being very close to the boundaries of the prior and (ii) the posterior for a $\chi_{\rm eff}=-0.8$ injection is much wider than the corresponding $\chi_{\rm eff}=0.8$ value, which does not show this unexpected behavior.

\subsection{q=7}
\label{subsec:q7}

\begin{figure*}
\includegraphics[width=1.0\textwidth]{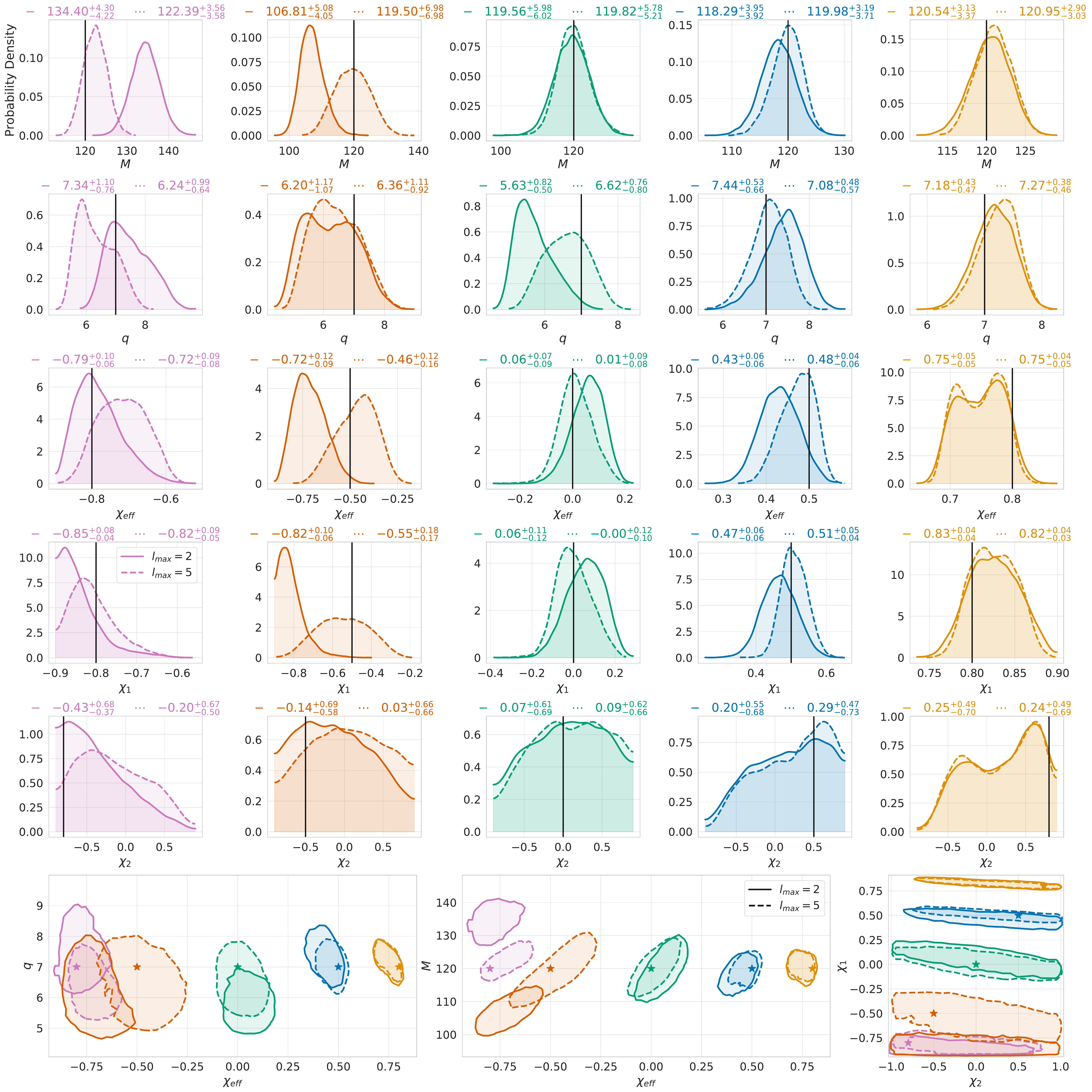}
\caption{\label{fig:q7} \textbf{Non-HM and HM runs for q=7 spin set, with SNR=30 and M=120}: The first five rows 
show the $M, q, \chi_{\rm eff}, \chi_{1z}, \chi_{2z}$ one-dimensional marginal distributions, where among this 
set of figures each column corresponds to a different synthetic source 
recovered with either all $\ell_{\rm max}=5$ modes (dashed line) or $\ell_{\rm max}=2$ modes (solid line). Our figures are organized such that the injected spin is systematically increased from left to right, where
the synthetic source runs 
are ID12 ($\chi_{\rm eff} = -.8$), ID13 ($\chi_{\rm eff} = -.5$), ID14 ($\chi_{\rm eff} = 0$), ID15 ($\chi_{\rm eff} = .5$), and ID16 ($\chi_{\rm eff} = .8$). In each figure's title, we report the median value and the 90\% confidence intervals of the marginalized 1D distribution for the $\ell_{\rm max}=2$ (left) and $\ell_{\rm max}=5$ (right) cases. A solid black vertical line denotes the true parameter value. The final bottom row corresponds to the joint distributions for $q$ vs $\chi_{\rm eff}$, M vs $\chi_{\rm eff}$, and $\chi_{1,z}$ vs $\chi_{2,z}$ for all five injections.
}
\end{figure*}

Finally, we analyze sources with  $q=7$. Figure \ref{fig:q7} shows the posterior distributions 
of the intrinsic parameters
for all the different spin configurations. The solid lines again represent runs that
were done with $\ell_{\rm max}=2$ modes, and the dashed lines represent runs that include all available $\ell_{\rm max}=5$.
As expected and consistent with the trend seen in the previous two subsections, we see substantial biases are often introduced in $M,q$ and $\chi_{\rm eff}$ if higher modes are omitted, especially for systems with large negative spin. Only the higher-mode model is able to make reliable parameter estimates, except for the large, positive spin configurations where a quadrupole-only model continues to do reasonably well. In some cases the the biased posterior doesn't even overlap with the true one,
which would be problematic for likelihood-reweighting techniques~\cite{2019arXiv190505477P}, which require similar posterior distributions.

Somewhat unexpectedly, however, is that the $\chi_1 = \chi_2 = 0.8$ system's posterior shows almost no effect from 
neglecting subdominant modes; any effect that is present is smaller than the corresponding equal-mass system with $\chi_1 = \chi_2 = -0.8$. We believe this can be explained by the orbital hangup effect~\cite{Campanelli2006c}, whereby given two otherwise identical systems the one with larger aligned spin will experience more orbits before merger. Consequently, the  $\chi_1 = \chi_2 = 0.8$ configuration will have more in-band cycles, and subdominant modes are known to be suppressed during the inspiral phase. We briefly elaborate on this effect in the conclusions.

\subsection{Effect of Network SNR on Biases}
\label{subsec:SNRs}
\begin{figure*}
\includegraphics[width=\textwidth,height=1.2\textwidth]{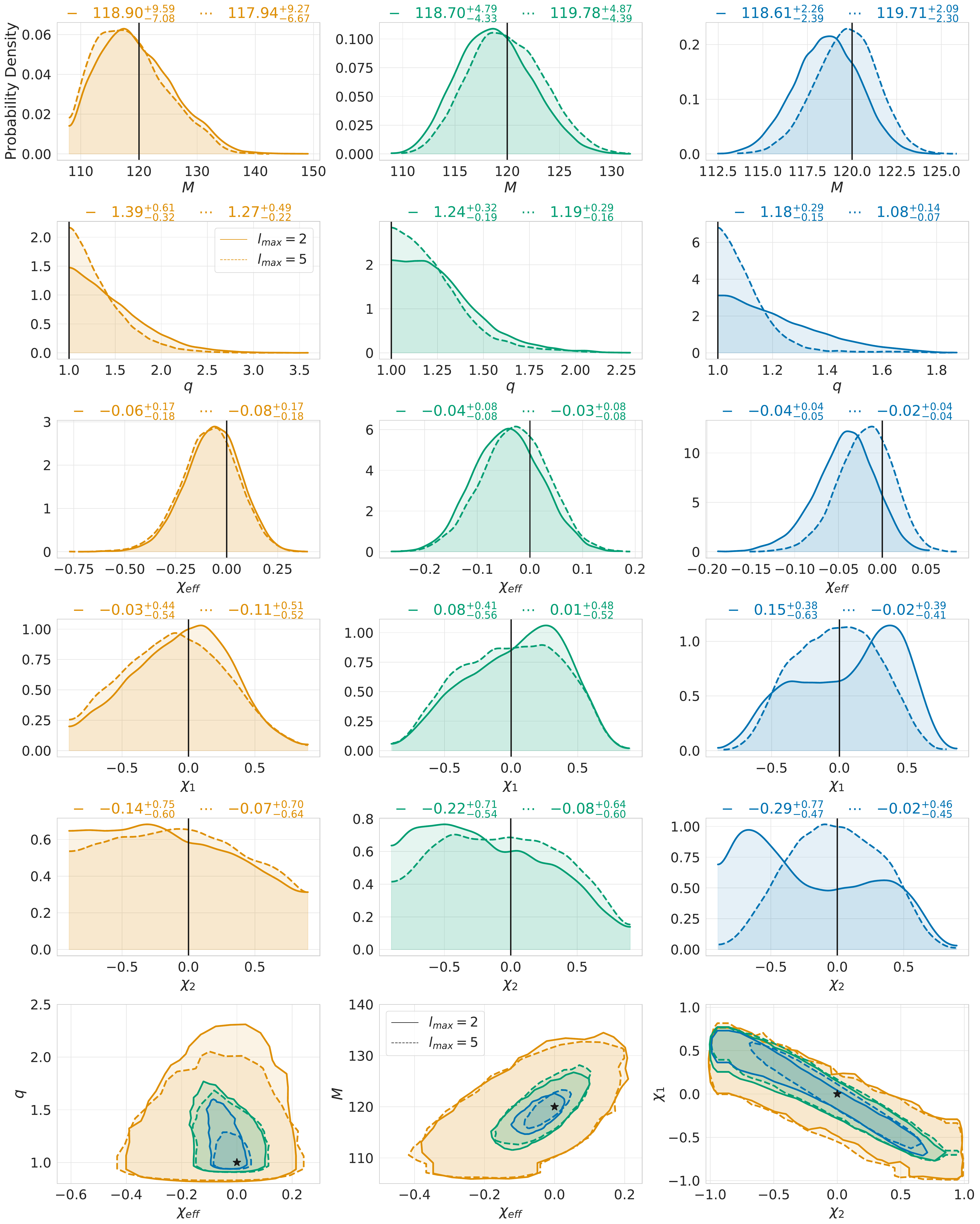}
\caption{\label{fig:SNR1}\textbf{Non-HM and HM runs for a q=1, M=120, and zero-spin source (ID4), for different SNRs}:
The first five rows 
show the $M, q, \chi_{\rm eff}, \chi_{1z}, \chi_{2z}$ one-dimensional marginal distributions, where among this 
set of figures each column corresponds to a different synthetic source 
recovered with either all $\ell_{\rm max}=5$ modes (dashed line) or $\ell_{\rm max}=2$ modes (solid line). Our figures are organized such that the signal's network SNR is systematically varied as 10 (orange), 30 (green), and 70 (blue), corresponding to the left, middle, and right columns, respectively. A solid black vertical line denotes the true parameter value. The final bottom row corresponds to the joint distributions for $q$ vs $\chi_{\rm eff}$, M vs $\chi_{\rm eff}$, and $\chi_{1,z}$ vs $\chi_{2,z}$ for all three injections.
}
\end{figure*}

\begin{figure*}
	\includegraphics[width=\textwidth,height=1.2\textwidth]{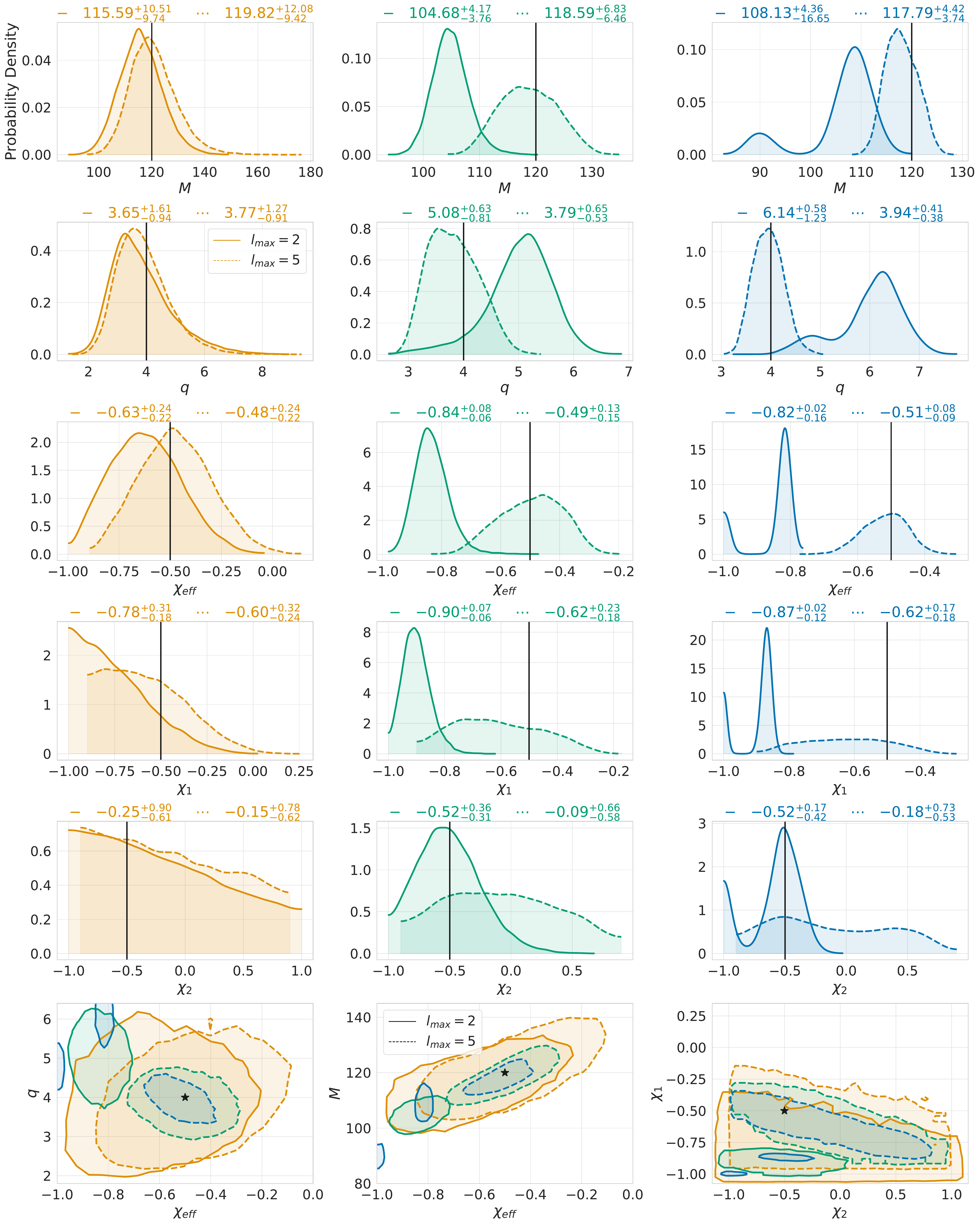}
	\caption{\label{fig:SNR2}\textbf{Non-HM and HM runs for a q=4, M=120, and $\chi_{\rm eff}$ = -0.5 source (ID8), for different SNRs}: The first five rows show the $M, q, \chi_{\rm eff}, \chi_{1z}, \chi_{2z}$ one-dimensional marginal distributions, where among this  set of figures each column corresponds to a different synthetic source recovered with either all $\ell_{\rm max}=5$ modes (dashed line) or $\ell_{\rm max}=2$ modes (solid line). Our figures are organized such that the signal's network SNR is systematically varied as 10 (orange), 30 (green), and 70 (blue), corresponding to the left, middle, and right columns, respectively. A solid black vertical line denotes the true parameter value. The final bottom row corresponds to the joint distributions for $q$ vs $\chi_{\rm eff}$, M vs $\chi_{\rm eff}$, and $\chi_{1,z}$ vs $\chi_{2,z}$ for all three injections.
    }
\end{figure*}

In the previous subsections, it was shown that a significant bias exists at SNR=30, even for the simplest systems. This
subsection is dedicated to investigating how the SNR affects the bias. Here we use all the different SNR runs from ID4
and ID8 in Table \ref{tab:SourceParameters}. Figures \ref{fig:SNR1} and \ref{fig:SNR2} show the posterior distributions for ID4 and ID8 respectively. As the SNR increases, the posteriors
become more precise for both the non-HM and HM results (i.e., the statistical errors get smaller). However, the HM results converge on the true parameters while
the non-HM results converge to a point offset from the true parameter (i.e., the systematic errors remain the same size and will dominate the statistical uncertainties). As GW detectors get more sensitive, the need for
HM will become paramount even for the simplest of events. More sensitive detectors will potentially bring into view
more exotic configurations at low SNRs which can also be problematic. For example, the weakest $q=4$, $\chi_{\rm eff} = -0.5$ system 
has noticeable bias. This could be anticipated by noting that the mismatch between $\ell_{\rm max}=5$ and $\ell_{\rm max}=2$ models at this injection value is $0.06989$ and so Eq.~\eqref{eq:indistinguishable} is not satisfied.

One particularly challenging configuration was the loudest $q=4$, $\chi_{\rm eff} = -0.5$ system shown in Fig.~\ref{fig:SNR2} (solid blue). In particular, the posterior recovered with the $\ell_{\rm max} = 2$ model shows evidence for a secondary peak widely separated from
the primary one. We checked this unexpected feature by directly comparing the values of the likelihood in a small neighborhood around both peaks. The presence of these two widely-separated peaks proved to be challenging for the current implementation of the ILE/RIFT algorithm, which uses a single interpolant of the log-likelihood surface. As a result, running this case took a significantly longer time while also achieving a comparatively lower accuracy, where the accuracy is quantified by the effective number of adaptive Monte Carlo samples. This case underscores that for high SNR events the omission of subdominant modes can introduce highly complex likelihood surfaces, and prove challenging to explore accurately. Within the RIFT framework, a recently implemented Gaussian Mixture Model sampler is expected to more efficiently sample from complicated likelihood surfaces. This case also demonstrates how incorrect models can accidentally yield good recovery of some parameters: the marginalized posterior for $\chi_2$ (solid blue curve) looks remarkably accurate around the primary peak despite the joint posterior (bottom right panel) being nowhere near the true value.

To quantify the bias between the non-HM and HM runs, we consider two commonly used measures of bias: (i) classifying the recovery of a particular parameter as biased if the injected parameter value is outside of the 90\% confidence region and (ii) the Jensen-Shannon divergence (JSD) between the different parameter distributions. Given two probability distributions $p(x)$ and $g(x)$, the JSD is defined as
\begin{equation}
  \label{eq:JSD}
  D_\mathrm{JS}(p \,|\, g) = \frac{1}{2} \Big(D_\mathrm{KL}(p\,|\,s) + D_\mathrm{KL}(g\,|\,s) \Big) \,,
\end{equation}
where $s = 1/2(p + g)$ and
\begin{equation}
  \label{eq:KLD}
  D_\mathrm{KL}(p\,|\,g) = \int p(x) \log_2\left( \frac{p(x)}{g(x)} \right) \mathrm{d} x \,,
\end{equation}
is the Kullback-Leibler divergence (KLD) between the distributions $p$ and $g$, measured in bits. For context, this is
the same calculation the LVC performed in \cite{LIGO-O2-Catalog} to quantify the agreement between different
models. When measured in bits, the JSD is bounded below by $0$. For a sense of scale, the KL divergence between two
one-dimensional Gaussians with identical standard deviations but differing means $\mu_1,\mu_2$ is
$(\mu_1-\mu_2)^2/2\sigma^2\ln 2$; inverting,  ${\rm JSD}=0.2$ corresponds to $\mu_1-\mu_2\simeq 0.5\sigma$. 

\begin{figure*}
\centering
\begin{minipage}[b]{0.49\textwidth}
\includegraphics[width=1.\textwidth]{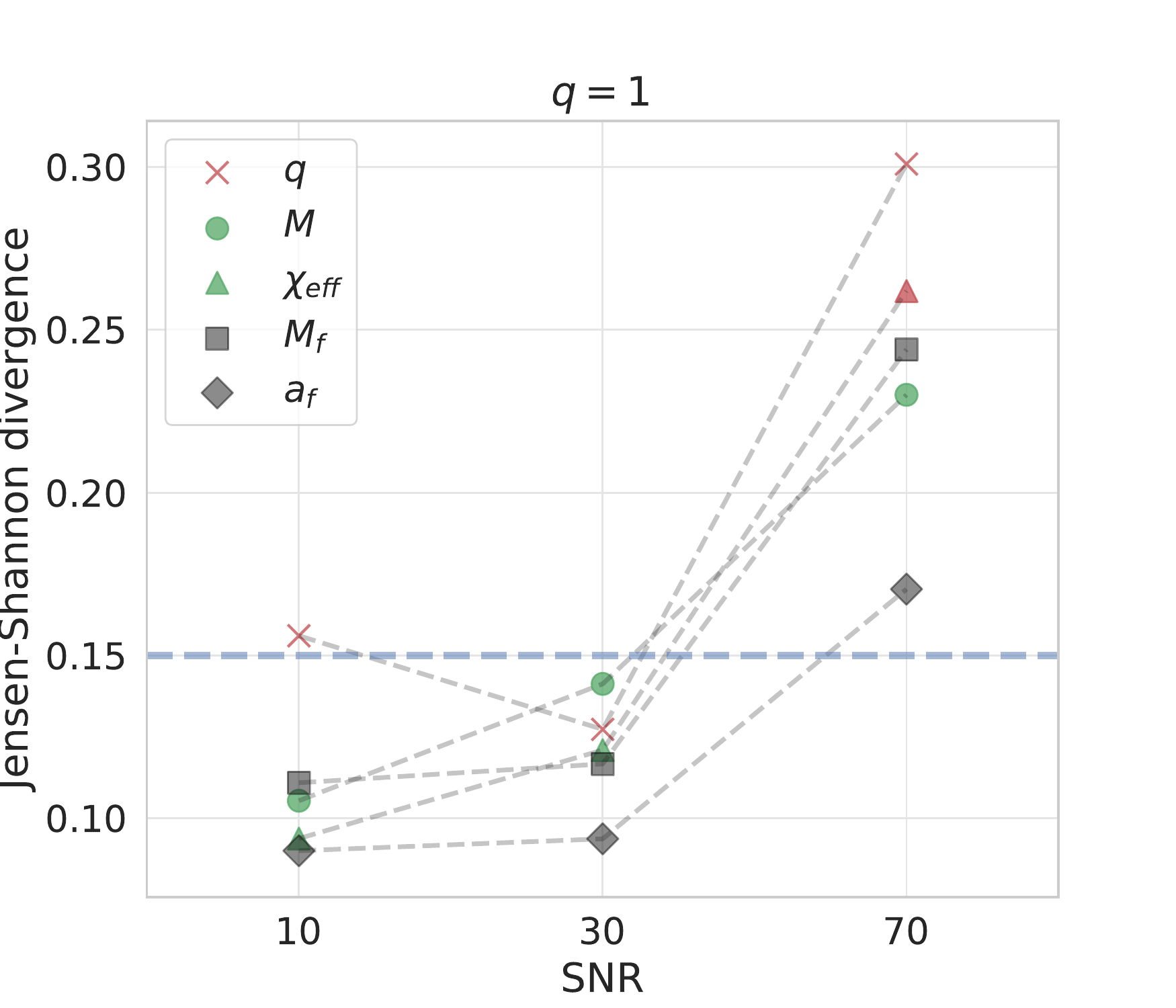}
%\caption{q = 1}
\end{minipage}
\hfill
\begin{minipage}[b]{0.49\textwidth}
\includegraphics[width=1.\textwidth]{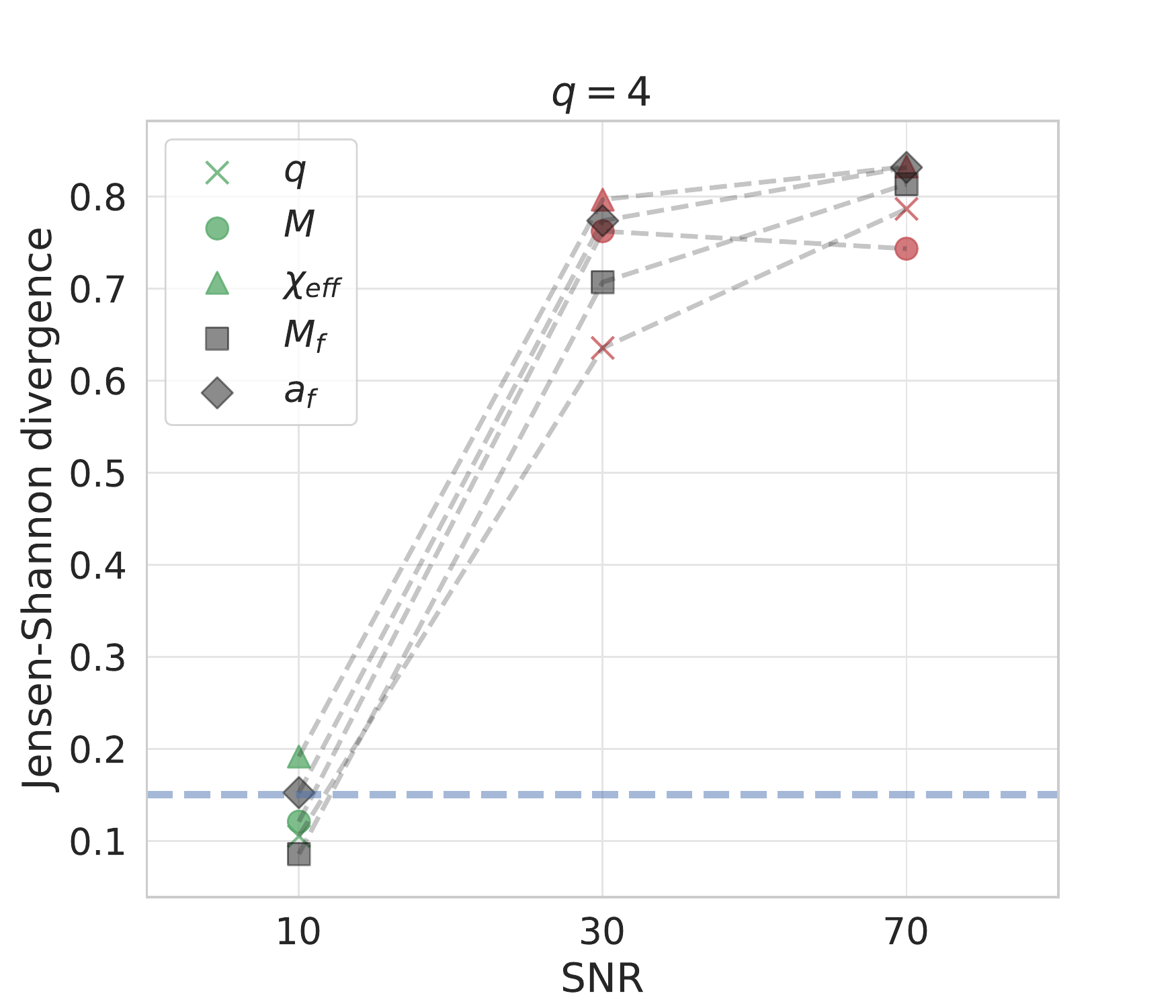}
%\caption{q = 4}
\end{minipage}
\caption{\label{fig:JSD}\textbf{The importance of higher modes for loud signals: bias vs SNR}: These panels show the JSD
  vs SNR for source ID4 (left panel) and ID8 (right panel). Different markers indicate which one-dimensional marginal
  distribution was used to evaluate the JSD, which are depicted in Figures (\ref{fig:SNR1}) and (\ref{fig:SNR2}) for ID4 and ID8, respectively. The dashed horizontal blue line demarcates a commonly used threshold for unacceptably large bias.  
  Markers colored in red indicate that the true value falls outside the 90\% credible interval region for the $\ell_{\rm max}=2$ case (significant bias in the recovered parameter value), while those colored in green indicate the opposite. For $\ell_{\rm max}=5$, the true value is almost always within the 90\% credible interval region except the parameter $q$ in the $q=1$ case, where the true value lies at the edge; despite not being in the he 90\% credible interval the marginalized distribution for $q$ obtains its maximum value at $q=1$ (cf. row 2 of Fig.~\ref{fig:SNR1}).
  Markers in gray indicate the JSD for the final remnant masses and spins.}
\end{figure*}

Figure \ref{fig:JSD} shows
the JSD vs SNR and the simple ``bias classifier" for both the ID4 and ID8 runs, respectively. 
Following the discussion in the LSC's recently published Gravitational-Wave Transient Catalog~\cite{LIGO-O2-Catalog} (cf. Appendix 2.B), we consider 
two marginalized posteriors to be sufficiently different (i.e. biased) if the JSD is greater than $\approx 0.15$.
This number corresponds to a SNR $\simeq 30$
for non-spinning, equal-mass binaries; SNR  $\simeq 10$ at $q=4$ and $\chi_{\rm 1z} = \chi_{\rm 2z} = -0.5$. Since subdominant modes become more important at larger mass ratios and more negative values of $\chi_{\rm eff}$, the quoted SNRs provide convenient lower bounds for similar systems. For example, we expect HMs will also affect the posterior for systems with SNRs $\geq 30$ and $q > 1$, $\chi_{\rm eff} \leq 0$ (similar to ID4); for systems  with SNRs $\geq 10$ and $q>4$, $\chi_{\rm eff} \leq -0.5$ (similar to ID8). 

%\clearpage
\section{Discussion} \label{sec:discussion}

\subsection{Effect of Different Spin Priors} \label{sec:spin_priors}
\begin{figure*}
\includegraphics[width=1.0\textwidth,height=1.2\textwidth]{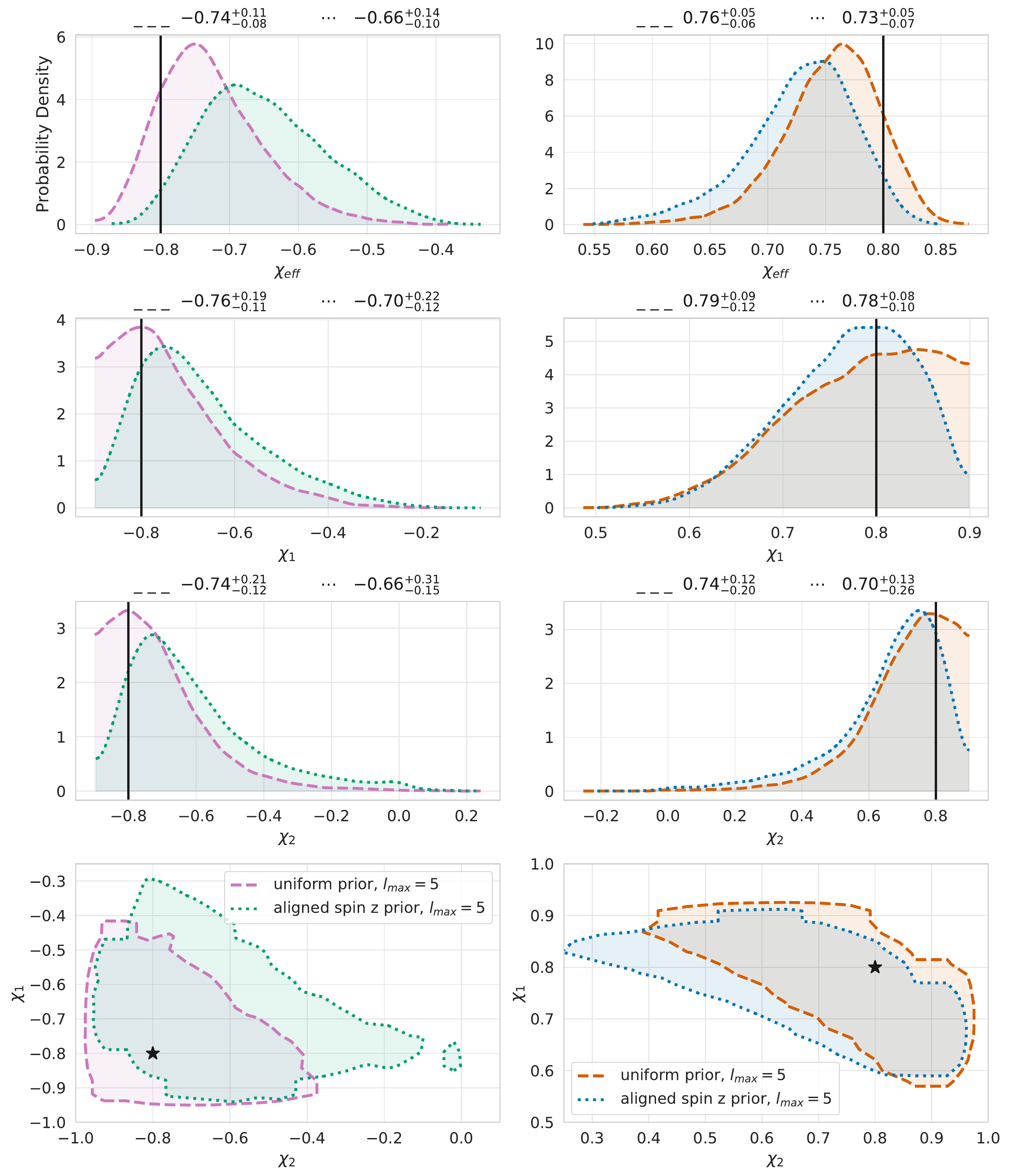}
\caption{\label{fig:priors1}\textbf{The effect of priors on spin measurability}: Individual and effective spin parameter recovery assuming two different priors, using synthetic datasets ID2 ($q=1$, $\chi_{\rm eff} = -.8$) and ID6 ($q=1$, $\chi_{\rm eff} = .8$) with SNR$=30$. The dashed curve represents the results using a prior that assumes uniform spin magnitudes in $\chi_{z}$ (P1; uniform prior), and the dotted curve represents the results using a prior that assumes uniform spin magnitudes in $\vec{\chi}$ (P2; aligned spin z prior). Despite the high value of SNR used here, we observe that the choice of prior has a significant influence on the recovered posteriors.} 
\end{figure*}

Besides the impact of sub-dominant modes, the ability to accurately measure the spin parameters is also influenced by the
choice in spin prior \cite{2018PhRvD..98d4028C}, which is not well-informed by astrophysical observations or source
population models.  In our study, we have used a prior which is uniform in  $\chi_{\rm z}$ (P1).
However, many of the LVC's analysis assume a prior that 
is uniform in spin magnitude, $|\vec{\chi}|$, and on the 2-sphere, which, for our non-precessing model, would induce a prior by projection of $\vec{\chi}$ along the orbital angular momentum vector (P2). When assuming this spin prior, the peak of the PDF of the individual component spins will strongly
favor zero. To see how these two significantly different priors affect the ability to measure the spins, we compare
posteriors for two runs ID2 and ID6 with SNR$=30$ assuming the two different priors. Figure \ref{fig:priors1} shows the
individual $\chi_{\rm * z}$ spins as well as the effective spin $\chi_{\rm eff}$ for each spin prior. Despite using a strong source, all spin parameters are
significantly perturbed by the prior choice, similar to results found in previous work \cite{2018PhRvD..98d4028C}.

\subsection{Consequences of biases for remnant properties and consistency tests}
\label{sec:spin}

Using the posterior distributions of the BBH system's component masses and spins one can compute the remnant mass, $M_f$, and spin, $a_f$ of the final (merged) black hole. The values of $(M_f,a_f)$ are interesting in their own right as they can be used to infer a population of astrophysical compact objects that formed through the merger of a BBH system. Another use of $(M_f,a_f)$ is to test the consistency of general relativity by predicting these remnant values found from (i) the post-merger portion of the signal which is described by a ringdown signal characterized entirely by $(M_f,a_f)$ and (ii) the inspiral portion of the signal where we compute the BBH system's component values and, using numerical relativity, arrive at an alternative estimate of 
$(M_f,a_f)$. If general relativity correctly describes the system's entire evolution, we should expect the remnant values found through each to be mutually consistent~\cite{TheLIGOScientific:2016src}. A closely related test uses the remnant values computed with the inspiral-only portion of the signal to infer the expected quasi-normal mode (QNM) of ringdown signal, and then comparing this predicted QNM spectrum with the QNMs estimated directly from the ringdown-only portion of the data~\cite{TheLIGOScientific:2016src}. A different, but related, set of tests of the no-hair theorem also benefit from the inclusion of both higher harmonics and as well as quasinormal mode overtones~\cite{Ota:2019bzl}.

All of these studies require accurate measurement of the system's remnant masses and spins. For example, unacceptably large bias in these quantities could provide misleading evidence for failed GR consistency tests, unless the quadrupole-only pre-merger and post-merger models make a serendipitously incorrect inference of the remnant properties (i.e., both models are incorrect but in a consistent manner).

In this subsection we explore bias in the remnant properties implied by the posterior distributions computed in Sec.~\ref{sec:bias} as the SNR increases. We compute the remnant mass and spin magnitude by evaluating the high-accuracy fitting formula provided by the surfinBH Python package~\cite{varma2019high} on the posteriors computed using $\ell_{\rm max}=5$ and $\ell_{\rm max}=2$ recovery waveform models.

\begin{figure*}
\includegraphics[width=\columnwidth,height=1.25\columnwidth]{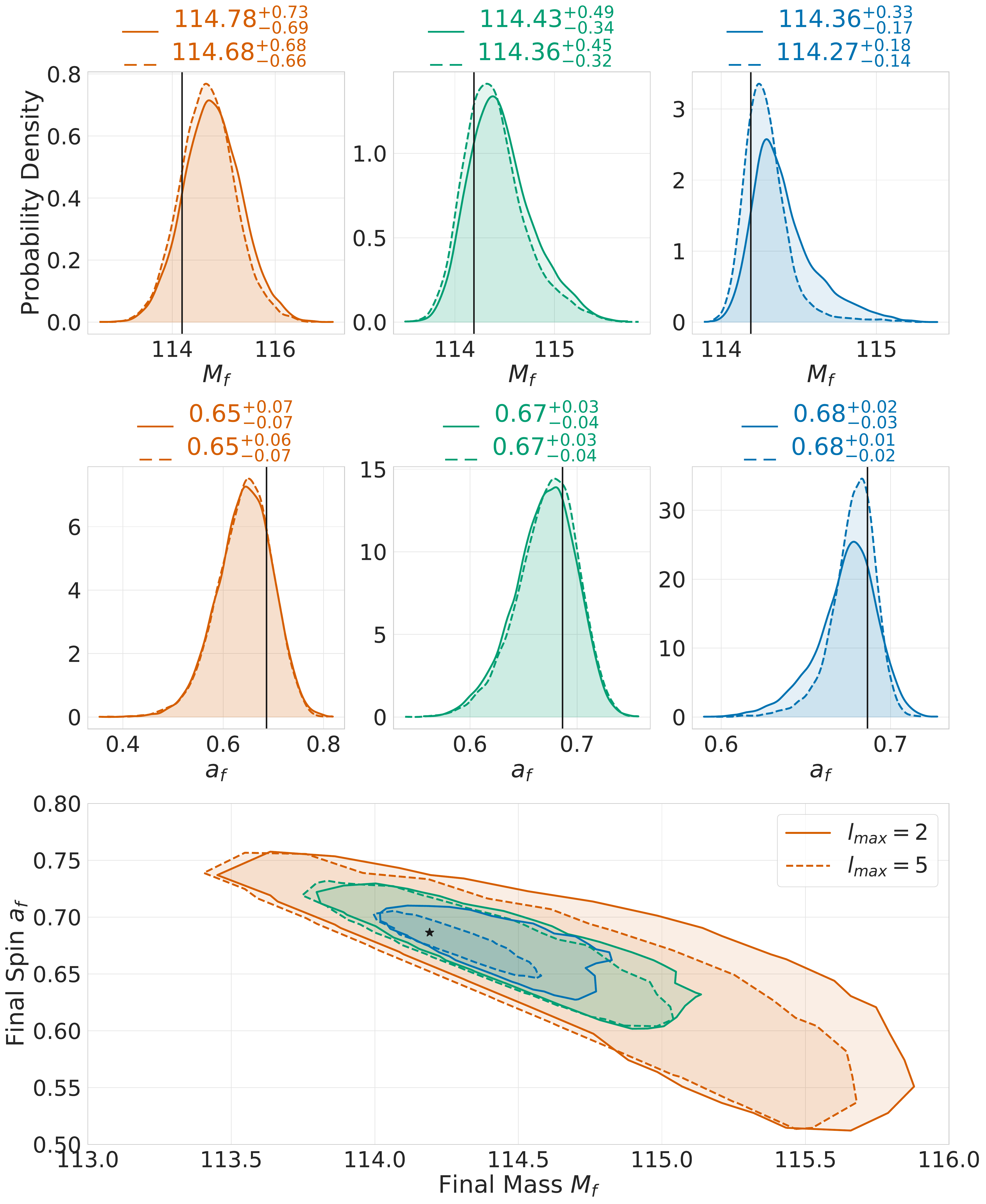}
\vrule
\vspace{1pt}
\includegraphics[width=\columnwidth,height=1.25\columnwidth]{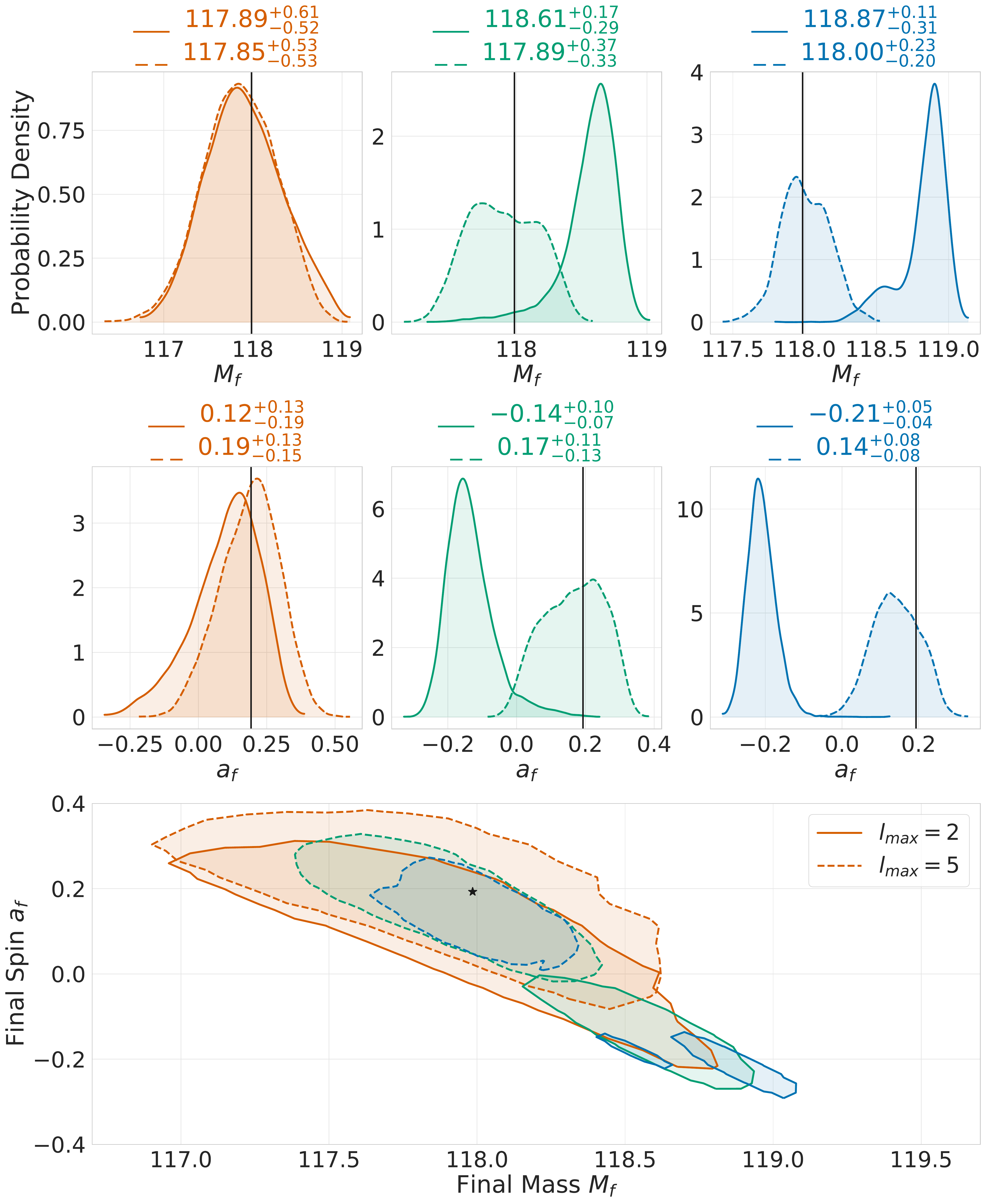}
\caption{\label{fig:remnants} 
\textbf{Effect of higher-order modes on remnant values and IMR consistency tests}: These panels show marginal distributions for remnant properties of the redshifted mass, $M_f$, and spin, $a_f$, for a non-spinning, $q=1$ source (ID4; left panels) and $\chi_{\rm eff} = -0.5$$, q=4$ source (ID8; right panels). Our figures are organized such that the signal's network SNR is systematically varied as 10 (orange), 30 (green), and 70 (blue), corresponding to the left, middle, and right columns of each panel.
}

\end{figure*}

As the first example, where we expect minimal bias, we consider the $q=1$, zero-spin source system (ID4) whose posterior distributions for SNRs$=\{10,30,70\}$ are reported in Fig.~\ref{fig:SNR1} from which we compute remnant posteriors in Fig.~\ref{fig:remnants} (left set of figures).  While the true remnant values are contained within all of the joint posteriors's 90\% credible region, we begin to see modest bias indicating impact from the higher-modes when the signal's strength reaches an SNR value of 70. This is quantified in Fig.~\ref{fig:JSD} which shows the Jensen-Shannon divergence 
for $M_f$ and $a_f$ are $0.24$ and $0.17$, respectively. For context, values above $0.15$ are typically considered to reflect non-negligible bias~\cite{LIGO-O2-Catalog}. At all values of the SNR, we find the $\ell_{\rm max}=5$ posterior more tightly constrains the true values.

Fig.~\ref{fig:remnants} also shows a similar sequence for the $q=4$, $\chi_{\rm eff}$ = -0.5 source (ID8) where now the true remnant values are no longer
contained within the 90\% credible intervals by SNR=30. As seen from Fig.~\ref{fig:JSD}, the JS divergence is already close
to, or greater than, $0.15$ at SNR=10. This suggests that higher modes are very important when estimating the remnant values from such systems, and neglecting them would incorrectly lead to a failure of the IMR consistency test for essentially any event we might conceivably observe similar to ID8.

\subsection{Consequences of biases on population reconstruction} \label{sec:astro}

% Astro
In a second and more qualitative example of the impact of parameter biases due to neglect of physics, we consider astrophysical inference for the mass, mass ratio and spin distribution
of coalescing BHs.   
 For example, consider an SNR=30, zero-spin BBH event with $q=4$. As illustrated by the green curves in Figure \ref{fig:q4},
inferences which neglect HMs would deduce negative effective spin (and a more extreme mass ratio). A single source with
definitively negative  $\chi_{\rm
  eff}$ would be interpreted as a strong indication for dynamical formation in samples of less than several hundred mergers.  Such biased inferences
for high-amplitude sources could thus be misinterpreted to support qualitatively different formation channels (e.g.,
dynamical formation) than supported by the true parameters, which are well-characterized by multimodal PE.      

More typically, parameter biases due to model incompleteness enter more insidiously into astrophysical inference, since
population inference relies on
combining information from multiple sources and since systematic biases impact all sources at a similar level.
Following \cite{gwastro-PopulationReconstruct-Parametric-Wysocki2018}, we estimate that parameter biases
$\Delta x = x^{\rm true} - x^{\rm median}$ will
be significant for a population of $N$ sources if the bias can be identified in the population mean by stacking
observations: in other words, if $\Delta x \gtrsim \sqrt{\sigma_{stat}^2+\sigma_{astro}^2}/\sqrt{N}$ where
$\sigma_{stat}$ and $\sigma_{astro}$ are the statistical error in $x$ and the width of the astrophysical distribution of
$x$, respectively. 
In terms of the JSD we anticipate that systematic differences in waveforms must produce a change in posteriors less than
${\rm JSD}=0.15/N$ to have no effect on population inference. 
Our examples show that even for zero-spin (but unequal-mass) binaries, inferences about the mass ratio, total mass, and effective spin in moderate-SNR sources can be
significantly biased by the lack of HM. If a population of unequal mass-ratio binaries exist and has a spin
distribution qualitatively similar to the seemingly low-spin BH population identified in O2, even inferences drawn from a handful of observations could be noticeably biased about BH masses and spins.

\subsection{A GW170729-like source}
\label{sec:o2-like}

While much of our focus has been on fiducial BBH systems, it is also interesting to consider  sources that are similar to events from the most recent observing run. In this subsection, we analyze a synthetic source that has parameters (cf. ID1 in Table \ref{tab:SourceParameters}) similar to GW170729, one of the more interesting events from O2. As mentioned in \cite{LIGO-O2-Catalog} and \cite{Kalaghatgi:2019log}, the SNR of GW170729 was $\thicksim12$. However, to better highlight the importance of HMs for this event, 
we instead consider a GW170729-like event located at a distance such that the SNR is $30$. For consistency with 
other synthetic events analyzed throughout this paper, we set $\chi_{\rm eff}=0.5$ as its true value, which is near the upper end of the 90\% credible interval t~\cite{chatziioannou2019properties}. Note that although $\chi_{\rm eff}=0.5$, we now have $\chi_{1z} \neq \chi_{2z}=0$. We continue using a uniform spin magnitude in $\chi_{\rm z}$ as our spin prior.

\begin{figure*}
	\includegraphics[width=\textwidth]{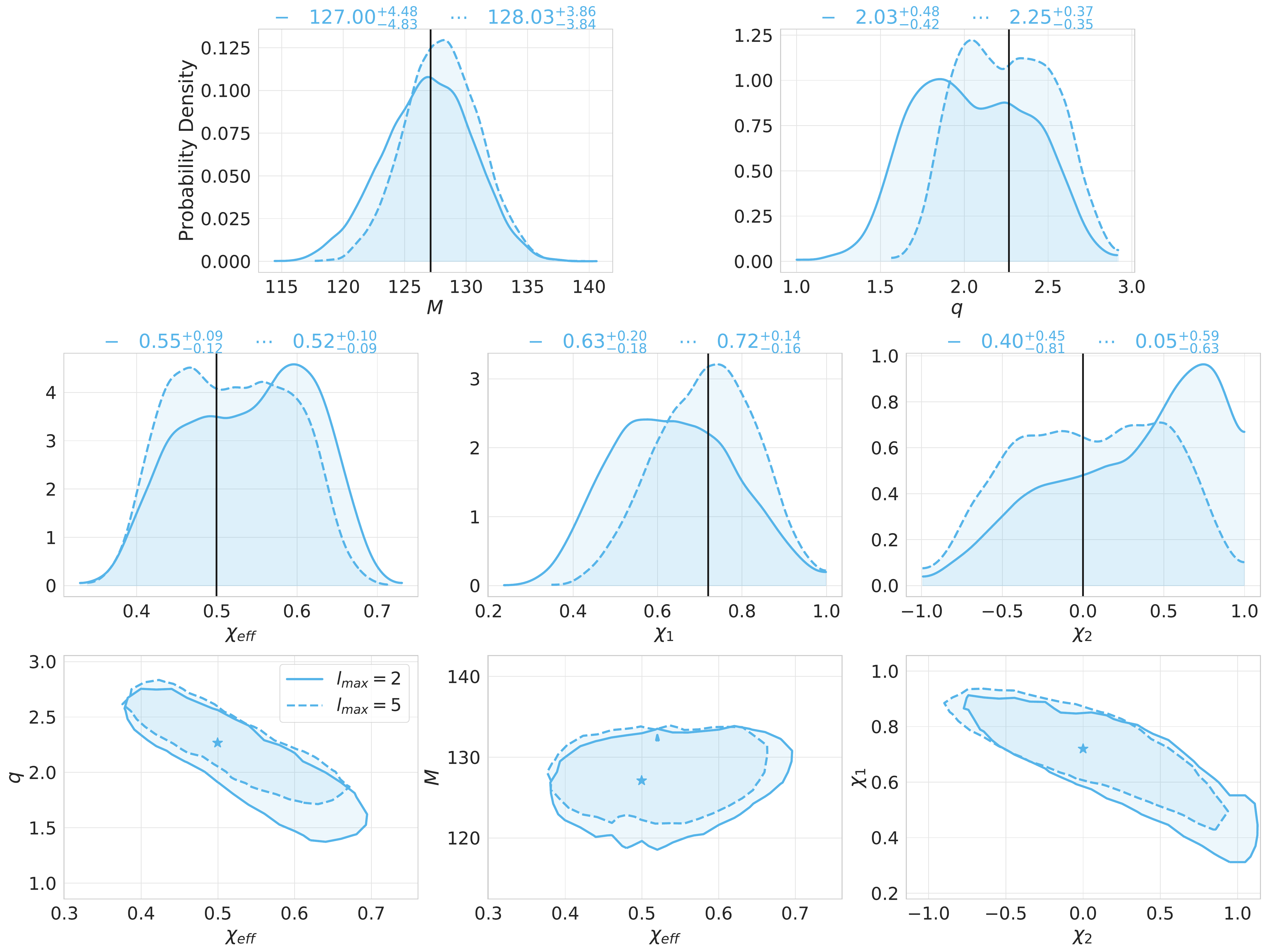}
	\caption{\label{fig:170729}\textbf{GW170729-like event}: Posterior plots for the ID1 run: $q$ = 2.267, $M$($M_{\odot}$) = 127.1, $\chi_{\rm 1z}$ = 0.72, $\chi_{\rm 2z}$ = 0.0, SNR$=30$. The solid and dashed lines represent the $\ell_{\rm max}=2$ and $\ell_{\rm max}=5$ runs respectively. When including HM, we are able to improve the recovery of individual spin components. We also see a significant shift in the $q$ and $\chi_{\rm eff}$ distributions.}
\end{figure*}

Figure \ref{fig:170729} shows the posterior distributions for the runs that include only $\ell_{\rm max}=2$ (solid lines) and include all the $\ell_{\rm max}=5$ (dashed lines). As with all the results in Section \ref{sec:bias}, we see a significant bias between the two runs in all the parameters. For example, we see that the $\ell_{\rm max}=5$ model does a much better job at recovering the individual spin components as well as placing somewhat tighter constraints on the spin of the larger BH, $\chi_{1z}$. Interestingly, we see a similar shift in $q$ and $\chi_{\rm eff}$ that was observed in a recent re-analysis of the actual GW170729 event~\cite{chatziioannou2019properties}. As our detectors continue to get more sensitive, we will increasingly see events with parameter and SNR  values similar to the synthetic source ID1 considered here.

\subsection{Comparison to previous works}
\label{sec:compare}

\begin{figure*}
	\includegraphics[width=\textwidth]{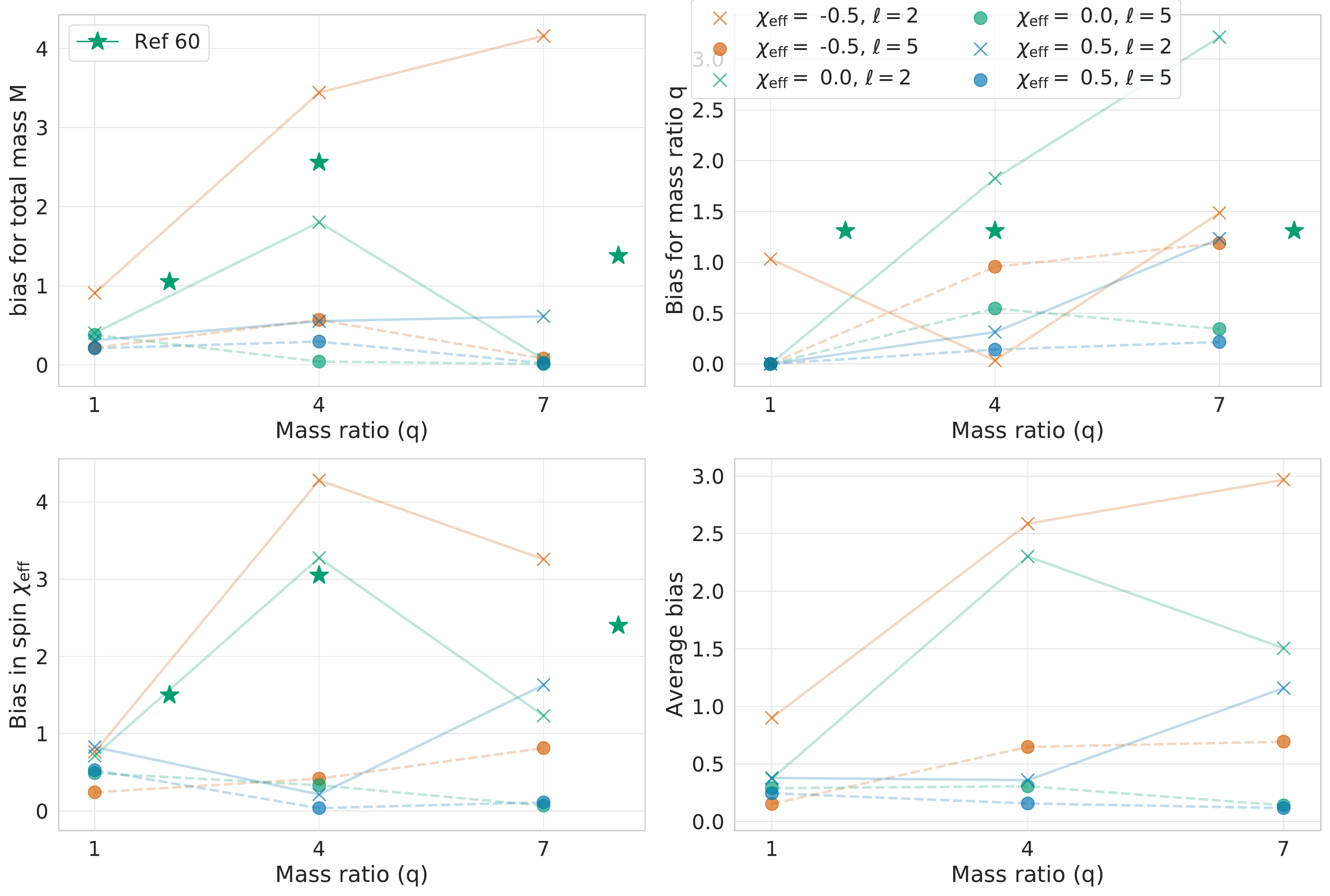}
	\caption{\label{fig:bias} Parameter biases, $\beta_{\lambda}$, for $\lambda = \{M, q, \chi_{\mathrm eff}\}$
	recovered with either all $\ell_{\rm max}=5$ modes (dashed line) or $\ell_{\rm max}=2$ modes (solid line).
	We consider different synthetic sources by varying $q=\{1,4,7\}$ and $\chi_{\mathrm eff} = \{-.5, 0, .5\}$
	while fixing SNR=30 and M=120. (Note: The marginalized posteriors for these systems are shown
	in Figures \ref{fig:q1}, \ref{fig:q4}, and \ref{fig:q7} which report the 90\% confidence interval values, instead of the 68\% values used to compute $\beta_{\lambda}$). To compare with Ref.~\cite{Kalaghatgi:2019log}, we
	also show the bias (green stars) for a similar set of nonspinning synthetic sources
	recovered with the quadrapole-only IMRPhenomD model. Broadly speaking, given the many differences
	in our setup, our findings are in general agreement with Ref.~\cite{Kalaghatgi:2019log}. Note that the apparent disagreement
	in $\beta_{q}$ appears to be due to our definition of the bias (see text). Finally, we also show the average bias,
	$(\beta_{M} + \beta_{q} + \beta_{\chi_{\mathrm eff}})/3$ in the bottom right panel. Here we clearly see general trends
	typically observed in our studies: subdominant modes are increasingly important as the value of the mass ratio increases 
        and/or effective spin decreases, and recovery models that include all
        modes reduces bias in all cases.
	}
\end{figure*}

Previous studies~\cite{Varma:2016dnf, bustillo2016impact, Capano:2013raa, Littenberg:2012uj,
Bustillo:2016gid, Brown:2012nn, Varma:2014, Graff:2015bba, Harry:2017weg,Kalaghatgi:2019log,LIGO-O1-PENR-Systematics,kumar2019constraining} have also considered the impact of subdominant modes on parameter estimation, 
and we have made qualitative comparisons to some of these works throughout our paper. 

In this subsection, we furnish a more quantitative
comparison by considering one commonly used measure of bias.
Instead of using the Jensen-Shannon divergence
to compare two marginalized posterior distributions, 
we now compute the bias,
\begin{align}
\beta_{\lambda} = \frac{\Delta \lambda}{\sigma_\lambda} \,,
\end{align}
as a ratio of the 
systematic error,
$\Delta \lambda = \left| \lambda_{\rm injected} - \lambda_{\rm recovered} \right|$,
to the $1\sigma$ statistical error
in the one-dimensional posterior, $\sigma_\lambda$.
The quantity $\beta_{\lambda}$ can be used to compare
with Varma et al.~\cite{Varma:2014,Varma:2016dnf} and
Kalaghatgi et al.~\cite{Kalaghatgi:2019log},
% and Littenberg et al.~\cite{Littenberg:2012uj}. 
We follow the choice of 
Refs.~\cite{Littenberg:2012uj}
where $\lambda_{\rm recovered}$ is taken to be
the maximum a posteriori (MAP) value. Note that  
Ref.~\cite{Kalaghatgi:2019log} instead defined
the recovered value to be the median value while Refs.~\cite{Varma:2014,Varma:2016dnf} used the
parameters that maximize the match, which is
similar to the maximum likelihood estimate.

We now summarize to what extent our results 
are consistent with previous ones. 
Broadly speaking, our findings are in 
agreement with both Kalaghatgi et al. and Varma et al., 
although there are some differences, which 
is to be expected. Indeed, our injected signals 
have larger SNRs, our
gravitational-wave recovery model is different, and
our setup uses a 
coherent Bayesian inference on the combined datasets from the
current three-detector network of observatories.

\subsubsection{Comparison to Varma et al.}

Refs.~\cite{Varma:2014,Varma:2016dnf} have used 
NR hybrids to map out where in the 
parameter space systematic errors from using quadrupole-only templates
dominate over the expected $1\sigma$ statistical errors. Such 
regions characterize where
neglecting subominant modes will lead to unacceptably large
errors in the parameter estimates. Statistical errors 
were estimated using Fisher information matrix approximations
with a single detector setup, while the value of $\lambda_{\rm recovered}$
was taken to be the best fit parameter values using a IMRPhenonD recovery model.
The injected signal's strength
was set to achieve a sky-averaged value of SNR=8 (corresponding
to an optimal orientation SNR of about 20), and they take a 
weighted average of the bias over a population of binaries with isotropic
orientations. Finally, while the effective spins of the injections they
consider are similar to ours, the individual spin components are different. 

Our main point of comparison is with Figure 1 of Ref.~\cite{Varma:2016dnf},
where the authors
identify where in the parameter space subdominant modes are 
important by considering where
$\beta_{\lambda}$ exceeds $1$. By this measure, in our study
subdominant modes are important for parameter estimation
for all of the cases shown in Fig.~\ref{fig:bias}
except $\chi_{\tt eff} = 0.5$ and $q \leq 4$. By comparison,
Varma et al find that nearly all of these cases show no bias; 
only $\chi_{\tt eff} = -0.5$ and large-mass ratio systems are 
require subdominant modes to be included in the model. 
As such, for heavy BBH systems, 
our results indicate 
that subdominant modes are required over a larger region
of the parameter space as compared to the general
conclusions of Ref.~\cite{Varma:2016dnf}. The most likely explanation for this
discrepancy is the different SNR values used in our studies. While typically
the largest SNR in any given detector is about $20$, our signal's network
SNR is $30$.

We also point out that all of the trends evident in
Fig.~1 of Ref.~\cite{Varma:2016dnf} have been confirmed in our
fully Bayesian, three-detector setup. Most interestingly that
at a fixed SNR the impact of subdominant modes will depend
strongly on $\chi_{\tt eff}$, with almost no bias observed for
large, positive spins. We return to this issue in the conclusions.

\subsubsection{Comparison to Kalaghatgi et al.}

A very recent study by Kalaghatgi et al.~\cite{Kalaghatgi:2019log}
used a two-detector Bayesian setup and studied the impact of subdominant modes
for non-spinning systems while systematically varying the inclination angle. 
In this study, NR hybrids are used as the signal template and 
a quadrupole-only IMRPhenomD recovery model is used. Indeed,
their choice of $M=100$ and SNR=25 makes their setup closely analogous to ours, which
facilitates direct comparison for non-spinning systems. We compare to their
set of runs where the injected signal's inclination is set to $60$ degrees, 
which is close to our value of $45$ degrees.

In Fig.~\ref{fig:bias} we plot (green star) the bias due to
omitting subdominant modes as 
reported in Ref.~\cite{Kalaghatgi:2019log}. These should be
compared with our non-spinning, $\ell_{\mathrm max} = 2$  (green circles; solid green line)
biases. The dependence of $\beta_M$ and $\beta_{\chi_{\tt eff}}$
with mass ratio is in broad agreement, with both results showing a similar up-down
pattern. Our smaller values of $\beta_M$ and $\beta_{\chi_{\tt eff}}$ indicate less error
due to neglecting subdominant modes, which
is somewhat surprising seeing as our network SNR is larger. This is most likely due to the 
fact that we inject and recover with the same NR surrogate model. 
Our values for $\beta_q$ appear to show disagreement, which is mostly due to 
differing choices for the recovered value. Indeed, since many of our posteriors 
in $q$ peak at $q=1$ the bias is 0, whereas the mean is offset from $1$. We have 
checked that when switching to the definition used in Kalaghatgi et al. 
our bias values are more consistent with values of about $1.4$, $1.4$ and $2.7$ 
at $q=1$, $q=4$, and $q=7$, respectively.

\subsection{Measuring individual black hole spins} \label{sec:spins}

It is well known that while individual spins are difficult to measure, the effective spin parameter, $\chi_{\rm eff}$, is much better constrained. A recent study~\cite{Purrer:2015nkh} systematically explored this question in the context of a single gravitational-wave detector by using the 
quadrapole-only SEOBNRv2 model~\cite{Purrer:2014,Taracchini:2013rva}. The general conclusion of this work (see Figures 1 and 4 of Ref.~\cite{Purrer:2015nkh}) is that individual spins are poorly constrained. For equal-mass systems, it was found that the spin measurements are constrained only by the Kerr limit and so only near-extremal spins can be constrained as the posterior will run up against the prior. Furthermore, as the mass ratio increases, the spin of the larger blackhole is better constrained while the smaller black hole's spin remains unconstrained. Finally, this general picture remains unchanged across a wide range of total masses, including the values we have focused on in our paper.

In this subsection, we revisit the results from Section \ref{sec:bias} but now briefly comment on our ability to measure the individual component spins using the full three-detector network with a our multi-mode recovery model. 

Unfortunately, as anticipated in Ref.~\cite{Purrer:2015nkh}, the inclusion of subdominant modes does not qualitatively change the 
situation. This is visually and quantitatively evident for equal mass (cf. Fig.~\ref{fig:q1}), $q=4$ (cf. Fig.~\ref{fig:q4}), and 
$q=7$ (cf. Fig.~\ref{fig:q7}) systems, all of which have a network SNR of 30. Here we see that while the inclusion of subdominant modes (dashed lines) dramatically reduces the bias in recovering  $\chi_{\rm eff}$, $\chi_{1}$, and $\chi_{2}$, the size of the 90\% confidence intervals (shown in the figure's title) are mostly unaffected. A similar conclusion can be reached by comparing the joint distributions for $\chi_{1}$ vs $\chi_{2}$ (bottom right panels in Figures \ref{fig:q1}, \ref{fig:q4}, and \ref{fig:q7}) recovered with $\ell_{\rm max} = 2$ and $\ell_{\rm max} = 5$ recovery models. 

Thus we conclude that, at least for the configurations considered here, including subdominant modes in our waveform recovery model will reduce bias in the both the effective spin and individual spin components, but does relatively little to better constrain them.

\section{Conclusions}
\label{sec:conclude}

In this work, using the recently-developed NRHybSur3dq8 model, we systematically investigate 
the importance of higher modes on the interpretation of gravitational wave signals
from coalescing binary
black hole systems. We have primarily focused on heavy systems with
masses and spins similar to the detector-frame masses of near-future
gravitational-wave observations while using current detector network sensitivities. 
Previous studies~\cite{Varma:2016dnf, bustillo2016impact, Capano:2013raa, Littenberg:2012uj,
Bustillo:2016gid, Brown:2012nn, Varma:2014, Graff:2015bba, Harry:2017weg,Kalaghatgi:2019log,LIGO-O1-PENR-Systematics,kumar2019constraining} have
also explored this question in various approximate contexts, either using a single detector, 
relying on Fisher information matrix approximations, or restricted to non-spinning BBH models. 
Here we perform coherent Bayesian inference on the combined datasets from the current three-detector network of observatories,
which is the same setup used in the recent analysis of gravitational wave observations~\cite{LIGO-O2-Catalog}.
We confirm many of the general expectations of previous works, while providing a more direct quantification of the bias 
within this realistic setup.

As expected, we find that higher
modes are very important for interpreting asymmetric binaries with $q>1$.  More surprisingly, we find noticeable
differences even when the injected signal mass ratio is $q=1$, when subdominant modes are expected to be suppressed (See Appendix \ref{app:zerospin} for a small follow up analysis). Also as expected, we find that the biases introduced by neglecting higher-modes are very important for $q>1$ and SNR$\ge 30$ \cite{Graff:2015bba,2017CQGra..34n4002O,Varma:2016dnf}.  However, in our examples we also find that inference without higher modes has a significant impact on
the interpretation of \emph{low-SNR} sources, particularly by influencing our knowledge of the binary's mass ratio.
General trends typically observed in our studies indicate that subdominant modes are increasingly 
important as the value of the mass ratio increases and/or effective spin decreases, and recovery models that include all modes reduces bias in all cases. Our work highlights the importance of subdominant modes for events similar to GW190412, an unequal mass BBH merger.

\begin{figure*}
\centering
\begin{minipage}[b]{0.49\textwidth}
\includegraphics[width=1.\textwidth]{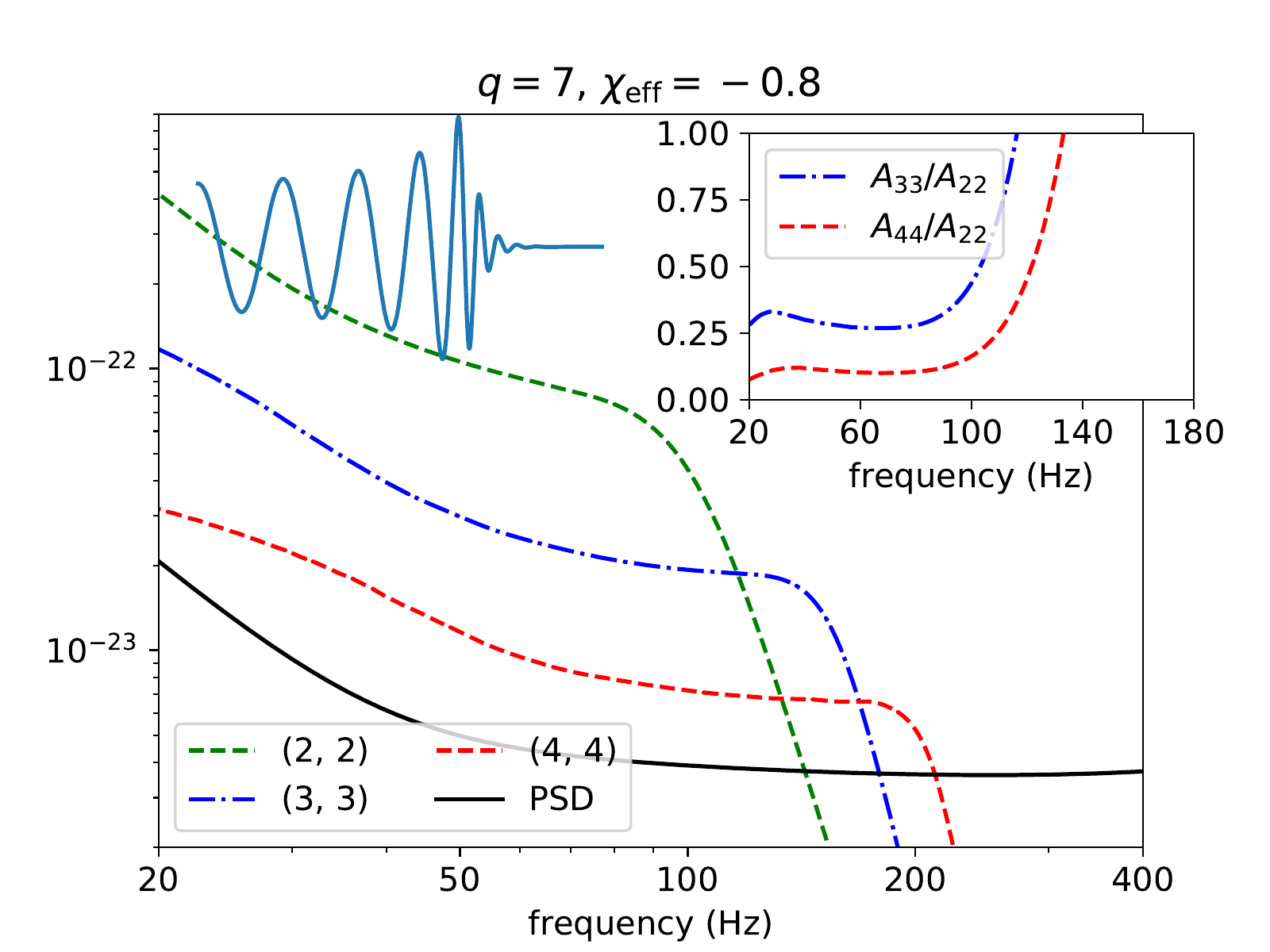}
%\caption{q = 1}
\end{minipage}
\hfill
\begin{minipage}[b]{0.49\textwidth}
\includegraphics[width=1.\textwidth]{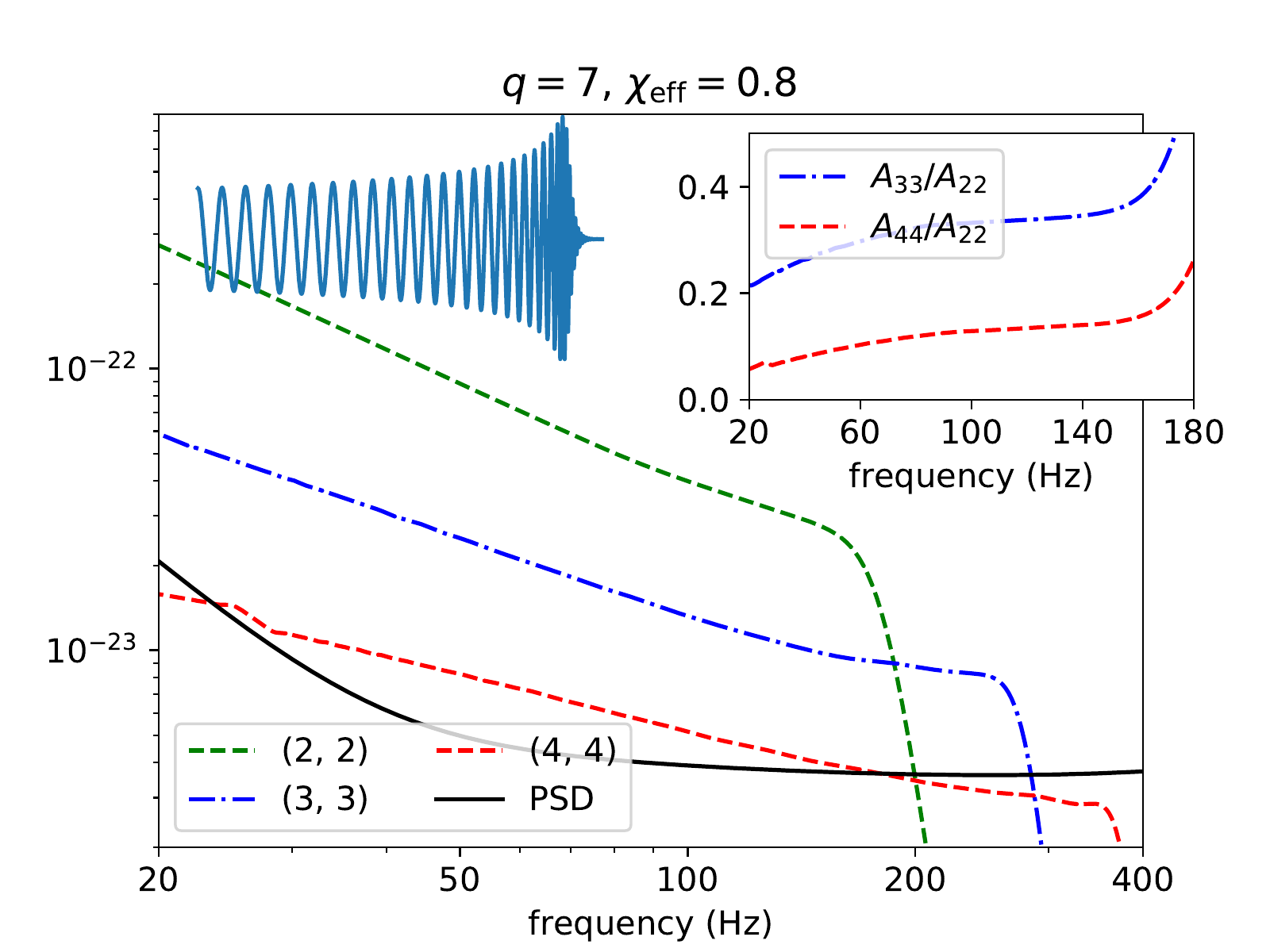}
%\caption{q = 4}
\end{minipage}
\caption{\label{fig:mode_power}These panels show the absolute value of the Fourier transform of the $(2,2)$, $(3,3)$, 
  and $(4,4)$ modes of the $q=7$ system with a spin of $\chi_{\rm eff} = -0.8$ (left; ID12) and
  $\chi_{\rm eff} = 0.8$ (right; ID16), and for reference we show the ZeroDetHighPower PSD. The inset figures show the ratio of the higher modes relative to the dominant mode. Because the $\chi_{\rm eff} = -0.8$ system merges at a lower orbital frequency, the subdominant modes contribute more to the overall SNR. This should be compared to the $\chi_{\rm eff} = 0.8$ system which merges at a higher orbital frequency, and consequently more of the inspiral portion of the waveform, which is dominated by the $(2,2)$ mode, contributes to the overall SNR. For illustrative purposes, a cartoon inset shows the time-domain signal starting from 20 Hz is of drastically different durations for these two systems. Note that the Fourier transformed signals were started from 3 Hz and tapered in order to avoid boundary effects. We also see that the waveform model has a small hybridization ``glitch" in the $(4,4)$ mode, which is likely due to post-Newtonian theory breaking down at high mass ratio and high spin; hybridization will be improved when higher order PN amplitude terms become available.
}
\end{figure*}

Consistent with previous work, we find that configurations with $M_z \simeq 120$ and large aligned spins have
almost no parameter bias~\cite{Varma:2016dnf} even at high SNRs. Such systems with large aligned spin exhibit the orbital hangup effect and have more in-band cycles. Given that the systems we have considered start in the late-inspiral regime, results from numerical relativity are most relevant toward quantifying the importance of this effect. For example, Table 3 from Ref.~\cite{Hannam2007c} shows that when starting from a fixed gravitational-wave frequency, the number of pre-merger orbits from an equal-mass, spin-aligned BBH system increases from about 5 to 9 as the effective spin parameter is varied from $0$ to $0.85$. 
Fig.~\ref{fig:mode_power} shows an example of this effect for the two most extreme cases we have considered in our study. The time-domain inset shows that the length of the signal increases as the spin becomes more positive, hence more of the SNR will be contained in the inspiral for systems with large, positive spin. The inspiral portion of the signal is known to be dominated by the $(2,2)$ mode's amplitude~\cite{bustillo2016impact}, which Fig.~\ref{fig:mode_power}'s insets show by comparing the relative amplitudes. We also see that near and after merger the higher modes quickly become larger in amplitude. Hence the impact of higher modes will be suppressed for longer signals, which seems to be why the orbital hangup effect serves to suppress the importance of higher modes. Other mechanisms by which more of the inspiral is in-band should similarly reduce the importance of higher harmonics. For instance, at a fixed SNR, the importance of subdominant modes for parameter estimation with systems with total masses lighter (heavier) than the fiducial value of $120$ considered here are expected to be less (more) important for parameter estimation.

In our examples, parameter inference of spinning BBH systems without higher modes are frequently biased. 
These consistent systematic biases may accumulate in population inference calculations, as described in
Sec.~\ref{sec:astro} (see also Ref.~\cite{gwastro-PopulationReconstruct-Parametric-Wysocki2018}).
We anticipate that any population inferences of
asymmetric, high-mass  black hole binaries will require significant attention to waveform systematics.  
Sec.~\ref{sec:spin} also demonstrated that neglecting higher-order modes in the analysis of GW observations
leads to biased estimates of the remnant object's mass and spin. For instance, Fig.~\ref{fig:remnants} 
shows that higher-modes provide significantly better constraints on the remnant values,
while the computed Jensen-Shannon divergence indicates 
a tension between the remnant mass and spin posteriors recovered by the $\ell_{\rm max}=2$
model and the true one (cf.~Fig.~\ref{fig:remnants}) over a range of SNRs and mass ratios.
As the remnant values feature prominently in
IMR consistency tests of general relativity, our study suggests that neglecting higher-modes 
could incorrectly trigger failed tests of GR, for example when carrying out
consistency tests between the strong-field merger and ringdown portions of the signal.
Despite the many benefits enumerated here, unfortunately, subdominant modes 
do not appear to improve our ability to resolve individual spin components, but they 
can reduce bias in their recovered values.

Finally, we have found that posteriors using an incomplete waveform model are often significantly
offset from the full-model posterior, typically towards (incorrectly) favoring lighter binary systems with more negative $\chi_{\rm eff}$ values. For example, a significant fraction of the probability for the $\ell_{\rm max}=5$ posterior
is not contained within the high-probability boundaries of the $\ell_{\rm max} = 2$ posterior. 
This suggests that it may be difficult to apply the
likelihood-reweighting techniques advocated in \cite{2019arXiv190505477P}, which require similar posterior
distributions in all binary intrinsic and extrinsic parameters for the two models being applied (i.e., a simplified $\ell_{\rm max} = 2$ model and a model including higher modes).

Given the large number of possible injection values one could consider, we
have restricted our attention to systems with  $M_z = 120 M_\odot$ and $\chi_{1z} = \chi_{2z}$,
while varying $\chi_{\rm eff}$, $q$ and the SNR. By relaxing these restrictions, 
future studies should explore the importance of subdominant modes with coherent Bayesian inference using 
the three-detector network of observatories. Within a restricted setup, 
previous studies have shown that, generally speaking, the bias due to omitting subdominant modes
increases at
higher total masses~\cite{bustillo2016impact,Varma:2014,Varma:2016dnf}. Given
that only the heaviest systems (e.g. GW170729) observed to date have a detector-frame
total mass near $M_z = 120 M_\odot$, our results provide a convenient 
upper bound on the greatest impact of subdominant modes for near-future
binary black hole observations. A more comprehensive survey using
our setup could be used to identify for which regions of the parameter space 
subdominant modes are important when considering total mass variations
(cf. Fig.~1 of Ref.~\cite{Varma:2016dnf}).

Looking ahead, we anticipate that aligned-spin IMR models including higher modes~\cite{london2018first,Cotesta:2018fcv,varma2019surrogate} will become standard in the analysis gravitational wave observations. Indeed, as shown here, the inclusion of subdominant modes will improve the interpretation of most events, and in some cases substantially so. 
A very recent study by Kalaghatgi et al.~\cite{Kalaghatgi:2019log}, using $M_z\simeq 120 M_\odot$, non-spinning BBH systems and an aligned-spin phenomenological recovery model IMRPhenomHM, has also concluded that higher modes significantly reduces bias. Using the most physically-complete models will also remove the need for ad hoc regions-of-validity that depend on both the source parameters as well as the scientific questions under consideration. However, to enable our model to fully encompass the range of likely events,
our models must also allow for generic precessing sources. Recent modeling of precessing binaries will allow for improved analysis of generic precessing sources~\cite{2019arXiv190509300V}. Indeed, as already indicated by Ref.~\cite{2018PhRvD..98b4019P}, we expect that many tests of general relativity could be biased unless they account for {\em both} higher modes and precession.

%\section{Acknowledgements}
\begin{acknowledgments}
We thank Gaurav Khanna for helpful discussions and providing technical 
assistance using the CARNiE cluster. We thank Chinmay Kalaghatgi and Juan Calderon Bustillo
for helpful comments on an earlier version of this manuscript, and the anonymous
referee for numerous suggestions.
ROS and JAL gratefully acknowledge NSF award PHY-1707965.~SEF is partially supported by NSF grant PHY-1806665,
and FHS is supported by NSF grant PHY-1806665 and the UMassD Physics Department.
L.E.K. acknowledges support from the Sherman Fairchild Foundation and NSF grant PHY-1606654 at Cornell.
The computational work of this project was performed on the CARNiE cluster
at UMassD, which is supported by the ONR/DURIP Grant No.\ N00014181255. 
SEF and FHS thank the Center for Scientific Computing \& Visualization Research (CSCVR)
for both its technical support and for its hospitality while part of this work was completed. 

\end{acknowledgments}

\appendix

\section{Follow up on the significance of higher modes for equal mass, zero spin, SNR$=10$ case}
\label{app:zerospin}

As pointed out in Sections \ref{subsec:SNRs} and \ref{sec:conclude}, there seems to be significant differences between the $\ell_{\rm max}=5$ and
$\ell_{\rm max}=2$ runs for the equal mass, zero spin, SNR$=10$ case, which runs contrary to several previous studies that had implied that HM would have minimal impact at low SNR for comparable-mass binaries. 

\begin{figure}[h]
\includegraphics[width=\columnwidth]{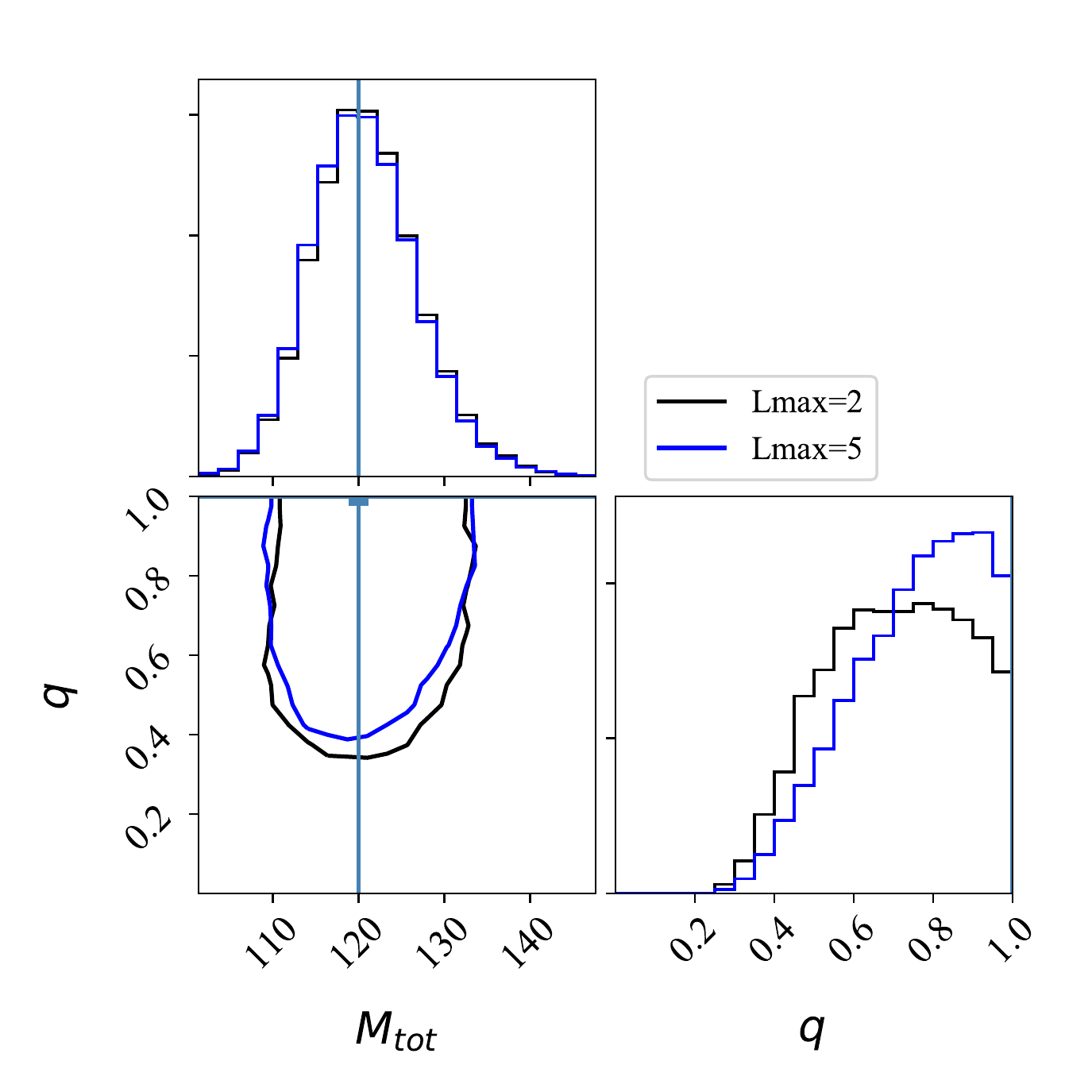}
\caption{\label{fig:zerospin}\textbf{Reanalysis of equal mass, zero spin, SNR$=10$}: This corner plot shows the reanalyses of a equal mass, zero spin, SNR$=10$ source using $\ell_{\rm max}=2$ (black) and $\ell_{\rm max}=5$ (blue) mode but only on a grid in mass parameters (i.e. assuming zero spin). As first shown in Figure \ref{fig:SNR1}, there are noticeable differences between the two different distributions.}
\end{figure}

To better understand our results, we perform a complementary analysis under the assumption of zero spin (i.e. lay out a grid only in $M_{\rm
  tot},q$), allowing us to directly evaluate the marginal likelihood versus the two remaining binary parameters. Figure
\ref{fig:zerospin} shows the results of both the $\ell_{\rm max}=5$ and $\ell_{\rm max}=2$ results. We continue to observe notable differences
between the two 
%analyses 
posteriors even when restricted to two dimensions (i.e. only mass parameters). It is certainly surprising to see any 
difference given that this is a low SNR, equal mass event. One possibility is that due to the broadness of the 
posterior in mass ratio, a significant fraction of the posterior needs to be evaluated at values of $q\gtrsim 2$ where 
higher modes begin to play an increasingly important role.

\clearpage 
%%%%%%%%%%%%%%%%%%%%%%%%%%%%%%%%%%%%%%%%%%%%%%%%%%%%%%%%%%%%%%%%%%%%%%%%%%%%%%%
\onecolumngrid
\section*{References}
\twocolumngrid
%%%%%%%%%%%%%%%%%%%%%%%%%%%%%%%%%%%%%%%%%%%%%%%%%%%%%%%%%%%%%%%%%%%%%%%%%%%%%%%
\bibliography{References}

\end{document}